\documentclass[%
 reprint,
superscriptaddress,
 amsmath,amssymb,
 aps,
 pra,
floatfix,
]{revtex4-2}
\usepackage{graphicx}
\usepackage{dcolumn}
\usepackage{bm}
\usepackage{braket}
\usepackage{amsfonts}
\usepackage{CJKutf8}
\usepackage[dvipsnames]{xcolor}
\usepackage{comment}

\usepackage[
    colorlinks=true,
    linkcolor=blue,
    citecolor=blue,
    urlcolor=blue
]{hyperref}
\usepackage[margin=1in]{geometry}
\newcommand{\h}[1]{\text{\footnotesize H}_{#1}}

\newcommand{\swap}{\text{\footnotesize SWAP}}
\newcommand{\cnott}[2]{\text{\footnotesize CNOT}_{#1,#2}}
\newcommand{\swapp}[2]{\text{\footnotesize SWAP}_{#1,#2}}
\newcommand{\cz}[2]{\text{\footnotesize CZ}_{#1,#2}}
\newcommand{\prep}{\text{\footnotesize PREP}} 
\newcommand{\select}{\text{\footnotesize SELECT}}
\newcommand{\measure}{\text{\footnotesize MEASURE}}
\DeclareMathOperator{\diag}{diag}

\newcommand{\cphase}
{\text{\footnotesize CPHASE}}

\usepackage[toc]{appendix}
\begin{document}

\title{Symmetry-preserving quantum compilation}
\author{Maryam Mudassar}
\email{maryammu@umd.edu}
\affiliation{Joint Center for Quantum Information and Computer Science, University of Maryland and NIST, College Park, MD, USA}
\author{Zhi-Yuan Wei (\begin{CJK*}{UTF8}{gbsn}魏志远\end{CJK*})}
\affiliation{Joint Center for Quantum Information and Computer Science, University of Maryland and NIST, College Park, MD, USA}
\author{Alexander Schuckert}
\affiliation{DIENS, École Normale Supérieure, PSL University, CNRS, INRIA, 45 rue d’Ulm, Paris 75005, France}
\affiliation{QuEra Computing Inc., 1380 Soldiers Field Road, Boston, MA 02135, USA}
\affiliation{QuEra Computing UK Ltd., Unit 2, Tech Foundry 2, Didcot, OX11 0DF}
\author{Michael J. Lawler}
\affiliation{Binghamton University, Binghamton, NY 13902, USA}
\affiliation{Cornell University, Ithaca, NY 14853, USA}
\author{Michael J. Gullans}
\affiliation{Joint Center for Quantum Information and Computer Science, University of Maryland and NIST, College Park, MD, USA}
\affiliation{QuEra Computing Inc., 1380 Soldiers Field Road, Boston, MA 02135, USA}
\author{Daniel Gottesman}
\affiliation{Joint Center for Quantum Information and Computer Science, University of Maryland and NIST, College Park, MD, USA}

\begin{abstract}
Fault-tolerant quantum simulation requires compiling symmetry-preserving time evolutions into discrete gate sets. Standard Clifford+T synthesis can break continuous symmetries even when the target evolution preserves them, causing compiled simulations to lose essential features of the original system. We develop a fault-tolerant compilation framework that preserves SU(2) symmetry exactly. We characterize the number-theoretic obstruction preventing symmetry-preserving unitary Clifford+T circuits from approximating arbitrary SU(2)-symmetric evolutions and circumvent it using measurement and feed-forward. The resulting synthesis confines compilation errors to the symmetry-preserving operator algebra rather than introducing symmetry-breaking perturbations.

We demonstrate the algorithm by simulating transport in the one-dimensional Heisenberg model. At a lower matched T-count, for instance, 54 T gates per two-site Trotter step, standard synthesis yields an incorrect transport exponent, whereas our symmetry-preserving compilation reproduces the expected exponent for the same expected T-count. If instead each approach is allocated whatever resources are necessary to estimate this exponent to the same accuracy, our construction reduces the T-count by approximately a factor of five. We further extend the framework to gauge symmetries in lattice gauge theories and to particle-number and spin symmetries in fermionic models relevant to quantum chemistry and the Fermi–Hubbard model.

These results show that symmetry-preserving fault-tolerant compilation can qualitatively change the effects of synthesis error while substantially reducing the resources required for quantum simulation
\end{abstract}

\maketitle
\section{Introduction}

Simulating the dynamics of quantum many-body systems is one of the most highly anticipated applications of quantum computing. First proposed by Feynman~\cite{Feynman1982} and placed on an algorithmic footing for local quantum systems by Lloyd~\cite{Lloyd1996}, the core motivation for quantum simulation stems from the exponential scaling of the Hilbert space, which renders exact classical simulation generically intractable as system size grows. Digital quantum simulation overcomes this barrier by mapping the target physical system onto a register of qubits and executing a sequence of quantum gates to approximate the continuous time-evolution operator, $U(t) = e^{-iHt}$. 

An important challenge in numerical simulations is to preserve the geometric structures and symmetries of the underlying physical system. In classical Hamiltonian dynamics, continuous symmetries give rise to conserved quantities through Noether’s theorem; rotational invariance, for example, implies conservation of angular momentum. Generic time-stepping methods may violate these invariants and produce unphysical drift. Structure-preserving methods are therefore often preferred: symplectic integrators preserve the Hamiltonian phase-space structure and typically exhibit bounded long-time energy errors—although they do not, in general, conserve the exact energy—while variational integrators can exactly preserve the discrete momentum maps associated with symmetries retained by the discretization \cite{GeometricIntegrationBook,WisdomSymplectic}. The Wisdom–Holman integrator used for long-time simulations of planetary systems is a prominent example \cite{WisdomSymplectic}. Analogous considerations arise in tensor-network simulations of quantum many-body systems. Global symmetries, such as \(U(1)\) particle-number conservation or \(SU(2)\) spin-rotation symmetry, can be imposed directly on the constituent tensors. This restricts the calculation to specified symmetry sectors, prevents symmetry-violating numerical errors, and produces block-sparse tensors that can substantially reduce memory requirements and contraction costs \cite{SinghTensorNetwork,SinghTensorNetwork2}.

A similar problem also features in quantum simulations, particularly when translating continuous dynamics to discrete dynamics.
The most ubiquitous approach to digital quantum simulation is Trotterization, which leverages Lie-Trotter-Suzuki product formulas to decompose the evolution of a complex global Hamiltonian into a sequence of discrete, implementable local operations~\cite{Childs2021Trotter}. However, these compilation methods do not necessarily preserve the symmetries of the underlying models. These continuous symmetries—such as $U(1)$ particle number conservation or $SU(2)$ rotational invariance—manifest as fundamental conservation laws that can constrain spectral degeneracies and strict selection rules for dynamical processes~\cite{LaBordeWilde2022}, and classify and protect topological phases. For a simulation to be physically meaningful, the simulated dynamics must not spuriously mix symmetry sectors.  

Translating these continuous, symmetry-constrained dynamics into discrete quantum circuits presents a major algorithmic challenge. Trotterization schemes that group interactions \textbf{globally} by Cartesian component (e.g., applying all Pauli $XX$ terms, then all $YY$ terms, and finally all $ZZ$ terms) can explicitly break global symmetries at generic finite Trotter step size. Consequently, the simulated quantum state unphysically leaks into incorrect symmetry sectors, corrupting the extraction of physical observables, correlation functions, and symmetry protected topological invariants. Recent work instead uses local symmetry, rather than commutativity alone, to guide the partition into Trotter blocks and thereby preserve physical structure while reducing product-formula error~\cite{NegishiYang2026LocalSymmetry}.

This symmetry-breaking problem is further exacerbated in the era of fault-tolerant quantum computing (FTQC). In fault-tolerant architectures based on a discrete logical gate set, arbitrary continuous rotations generally must be synthesized, most commonly using Clifford+$T$. Because standard synthesis algorithms (such as the Solovay--Kitaev or \textit{gridsynth} algorithms)~\cite{RossSelinger2016} are inherently symmetry-agnostic, approximating an arbitrary continuous rotation introduces approximation errors. These synthesis errors act as coherent noise that can break the underlying continuous symmetries of the physical model over deep circuits.

In this work, we develop fault-tolerant compilation methods that preserve physical symmetries exactly, even when the target evolution can only be approximated. Using the \(SU(2)\)-symmetric one-dimensional Heisenberg model as a testbed, we express the two-site exchange evolution in terms of fractional SWAP gates and characterize the number-theoretic obstruction to approximating arbitrary exchange angles with unitary Clifford+\(T\) circuits while remaining within the \(SU(2)\)-symmetric operator algebra. We overcome this obstruction by allowing ancilla measurements and classical feed-forward. In particular, we introduce a linear-combination-of-unitaries (LCU) protocol that directly synthesizes the controlled-phase operation required in the singlet--triplet basis. We compare this construction with conventional symmetry-agnostic synthesis and with the repeat-until-success protocol of Bocharov, Roetteler, and Svore (BRS)~\cite{Bocharov2015RUS}, which can also be used within our decomposition to produce symmetry-preserving exchange gates. In numerical simulations of spin transport, the symmetry-preserving methods recover the expected KPZ transport exponent at resource budgets for which conventional synthesis gives an inaccurate result. Over the parameter range studied, our LCU construction also achieves lower expected \(T\)-count and higher conditional fidelity than the BRS-based construction. Finally, we explain how the same operator-algebraic framework extends beyond the Heisenberg model to local gauge symmetries and to particle-number and spin symmetries in fermionic models relevant to quantum chemistry and the Fermi--Hubbard model.

\textit{Outline:} Section~\ref{section: background} reviews the Heisenberg model and product-formula simulation. Section~\ref{sec:swap-roots} describes the symmetry-adapted decomposition of exchange evolution into fractional SWAP gates and identifies the angles that can be implemented exactly over Clifford+\(T\). Section~\ref{sec:fault-tolerant} establishes the number-theoretic obstruction to approximating arbitrary \(SU(2)\)-symmetric exchange rotations using symmetry-preserving unitary Clifford+\(T\) circuits. Section~\ref{sec:lcu} introduces our measurement-assisted LCU construction and compares it with conventional synthesis and the BRS protocol. Section~\ref{sec:numerical-benchmarks} benchmarks these methods using spin transport in the one-dimensional Heisenberg model, focusing on the extracted KPZ exponent, fault-tolerant resource costs, state fidelity, and leakage between symmetry sectors. Section~\ref{sec:extensions} develops extensions to other symmetry-constrained systems, including lattice gauge theories and fermionic models.
\section{Background}\label{section: background}
\subsection{The Heisenberg Model}\label{section: Heisenberg}

The Heisenberg model serves as a paradigmatic system for exploring quantum magnetism, entanglement dynamics, and quantum phase transitions \cite{Piroli2016,Manousakis1991, Amico2008}. Notably, the one-dimensional (1D) isotropic Heisenberg model is exactly solvable via analytical methods~\cite{Bethe1931}. This exact solvability makes it a useful theoretical tool for testing new numerical metrics and quantum algorithms \cite{White1992, Childs2018}.

The general Hamiltonian for the isotropic Heisenberg model describes nearest-neighbor exchange interactions and can be written as:
\begin{equation}
    H = J_{ex} \sum_{\langle i,j \rangle} \left( X_i X_j + Y_i Y_j + Z_i Z_j \right) = J_{ex} \sum_{\langle i,j \rangle} \vec{\sigma}_i \cdot \vec{\sigma}_j
\end{equation}
where $\vec{\sigma}_i = (X_i, Y_i, Z_i)$ is the vector of Pauli operators acting on site $i$, $J_{\mathrm{ex}}$ is the exchange coupling constant (with $J_{\mathrm{ex}} > 0$ yielding antiferromagnetic and $J_{\mathrm{ex}} < 0$ ferromagnetic order), and $\langle i,j \rangle$ denotes the sum over nearest-neighbor pairs. We set $J_{ex}=1$ in the remainder of the paper. 

Restricting the geometry to a one-dimensional chain, the Hamiltonian simplifies to:
\begin{equation}
    H_{\text{1D}} = J_{ex} \sum_{i=1}^{N-1} \vec{\sigma}_i \cdot \vec{\sigma}_{i+1}
\end{equation}

A defining feature of the isotropic Heisenberg model is its global $SU(2)$ rotational symmetry. Let us define the total spin operators $J_\alpha = \frac{1}{2} \sum_i \sigma_i^\alpha$ for $\alpha \in \{x,y,z\}$, and the total spin squared operator $J^2 = J_x^2 + J_y^2 + J_z^2$. Because the Hamiltonian is composed entirely of rotationally invariant scalar products, it commutes with all generators of the $SU(2)$ group as well as the Casimir invariant:
\begin{equation}
    [H, J_\alpha] = 0, \qquad [H, J^2] = 0
\end{equation}
Consequently, the time-evolution operator $U(t) = e^{-iHt}$ also respects these symmetries, ensuring that the total spin of the system is conserved throughout its evolution.

Finally, it is often mathematically and algorithmically insightful to express $H_{\text{1D}}$ in terms of the two-qubit exchange (or SWAP) operator, $P_{i,j}$. Using the identity $\vec{\sigma}_i \cdot \vec{\sigma}_j = 2P_{i,j} - I$, the Hamiltonian takes the form:
\begin{equation}
    H_{\text{1D}} = J_{\mathrm{ex}} \sum_{i=1}^{N-1} \left( 2P_{i,i+1} - I \right)
    \label{eq:swap_heisenberg}
\end{equation}
Because the total spin squared $J^2$ is fully symmetric under particle exchange, this $\swap$-based representation makes it manifestly clear why $[H, J^2] = 0$.

\subsection{Overview of Trotterization}

The simulation of quantum dynamics requires implementing the time evolution operator $U(t) = e^{-iHt}$ for a given system Hamiltonian $H$ (assuming $\hbar = 1$). In most practical quantum systems, the Hamiltonian can be expressed as a sum of local, manageable sub-Hamiltonians:
\begin{equation}
H = \sum_{k=1}^{m} H_k
\end{equation}
where the individual terms $H_k$ are relatively simple to simulate or implement as discrete quantum operations. However, because these individual sub-Hamiltonians generally do not commute with one another ($[H_i, H_j] \neq 0$ for some pairs $i \neq j$), the exponential of their sum does not equal the product of their exponentials.

To circumvent this non-commutativity, we employ the Lie-Trotter product formula, commonly referred to as Trotterization. This technique approximates the total continuous time evolution by slicing the time parameter $t$ into $n$ small, discrete intervals (or ``Trotter steps'') of duration $\Delta t = t/n$. The first-order Trotter approximation is given by:
\begin{equation}
e^{-iHt} = e^{-i \sum_{k=1}^m H_k t} = \lim_{n \to \infty} \left( \prod_{k=1}^{m} e^{-iH_k t/n} \right)^n
\end{equation}

For a finite number of steps $n$, the first-order approximation introduces an algorithmic error known as the Trotter error. The approximate time evolution operator $\tilde{U}(t)$ takes the form:
\begin{equation}
\tilde{U}(t) = \left( \prod_{k=1}^{m} e^{-iH_k t/n} \right)^n = e^{-iHt} + \mathcal{O}\left(\frac{t^2}{n}\right)
\end{equation}
By choosing $n$ to be sufficiently large, the Trotter error $\mathcal{O}(t^2/n)$, which is dictated by the commutators of the sub-Hamiltonians, can be suppressed to fall within any desired error tolerance $\epsilon$. Higher-order Suzuki--Trotter formulas can likewise be employed to reduce the approximation error without altering the basic structure of the simulation. For example, the symmetric second-order formula,
\begin{equation}
e^{-iH\Delta t}
=
\left(\prod_{k=1}^{m} e^{-iH_k\Delta t/2}\right)
\left(\prod_{k=m}^{1} e^{-iH_k\Delta t/2}\right)
+\mathcal{O}(\Delta t^3),
\end{equation}
leads to a global error of $\mathcal{O}(t^3/n^2)$, while higher-order recursive formulas provide further improvements. Importantly, the symmetry arguments developed below apply independently of the order of the product formula. If a symmetry operator $S$ is conserved by every sub-Hamiltonian, such that $[H_k,S]=0$ for all $k$, then each local evolution $e^{-iH_k\tau}$ commutes with $S$. For instance, the Heisenberg model in Section \ref{section: Heisenberg} commutes with $S=J^2, \bm{J}$, where $\bm{J=(J_x,J_y,J_z)}$ is spin angular momentum vector, and $J^2 = \bm{J}\cdot \bm{J}$.  For the bond-based decomposition of the isotropic Heisenberg model, \(H_k=J_{\mathrm{ex}}\boldsymbol{\sigma}_k\cdot\boldsymbol{\sigma}_{k+1}\), every \(H_k\) commutes with \(J_x\), \(J_y\), \(J_z\), and \(J^2\). Consequently, every product-formula step constructed from these bond evolutions preserves the global \(SU(2)\) symmetry exactly. including for higher-order decompositions.

\section{Roots of SWAP Gates and Symmetries}
\label{sec:swap-roots}

Exchange gates provide natural symmetry-preserving primitives for the Heisenberg model. For $N$ spin-$1/2$ particles, the total-spin Casimir can be written in terms of pairwise SWAP operators $P_{i,j}$ as
\begin{equation}
    J^2 = \frac{3N}{4}I
    + \frac{1}{2}\sum_{i<j}(2P_{i,j}-I).
\end{equation}
Moreover,
\begin{equation}
    P_{i,j}
    = \frac{1}{2}
      \left(I+\vec{\sigma}_i\cdot\vec{\sigma}_j\right)
\end{equation}
is rotationally invariant and commutes with each global spin generator $J_\alpha$, $\alpha\in\{x,y,z\}$. Hence, exchange gates preserve the global $SU(2)$ symmetry.

Using Eq.~\ref{eq:swap_heisenberg}, the evolution of a nearest-neighbor pair is
\begin{equation}
    U_{i,i+1}(\delta t)
    = e^{-i\delta t
        (\vec{\sigma}_i\cdot\vec{\sigma}_{i+1})}
    = e^{i\delta t}
      e^{-2i\delta t P_{i,i+1}}.
\end{equation}
Thus, different Trotter step sizes correspond to different fractional powers of SWAP, a standard resource in exchange-based quantum computation~\cite{DiVincenzo2000Exchange}.

\begin{figure}
    \centering
    \includegraphics[
        trim=5cm 20cm 5cm 3cm,
        clip,
        width=0.5\textwidth
    ]{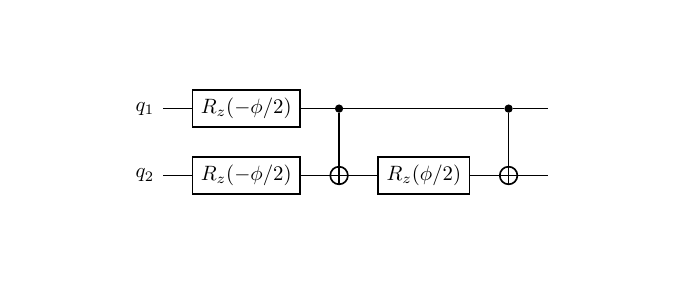}
    \caption{Decomposition of the $\cphase(\phi)$ gate.}
    \label{fig:cphase}
\end{figure}

These powers can be reduced to elementary one- and two-qubit gates using the standard construction of Ref.~\cite{Barenco1995Elementary}. In particular,
\begin{equation}
\begin{split}
    P_{1,2}
    &= \cnott{1}{2}\,\cnott{2}{1}\,\cnott{1}{2} \\
    &= \cnott{1}{2}\,\h{1}\,\cz{1}{2}\,
       \h{1}\,\cnott{1}{2}.
\end{split}
\end{equation}
More generally, for $\delta t=\pi/2^{n+2}$,
\begin{equation}
\begin{split}
    (P_{i,i+1})^{1/2^n}
    &= \cnott{i}{i+1}\,\h{i}\,
       \cphase(4\delta t) \\
    &\quad {}\times
       \h{i}\,\cnott{i}{i+1},
\end{split}
\end{equation}
where
\begin{equation}
    \cphase(4\delta t)
    = \operatorname{diag}
      (1,1,1,e^{i4\delta t}).
\end{equation}
Its decomposition into CNOT gates and local $Z$ rotations, up to a global phase, is shown in Fig.~\ref{fig:cphase}.

Within this family, SWAP, $\sqrt{\mathrm{SWAP}}$, and
$\mathrm{SWAP}^{1/4}$ are exactly representable over
Clifford+$T$. Their corresponding Trotter angles are listed
in Table~\ref{tab:exact-swap-roots}; other angles require
approximation. Moreover, for $\delta t=\pi/2^{n+2}$, the resulting
$\cphase(\pi/2^n)$ gate lies at level $n+2$ of the
Clifford hierarchy and can be implemented fault tolerantly
using the corresponding magic-state injection protocol. In the remainder of this work, however, we restrict the underlying unitary gate set to Clifford+\(T\)

\begin{table}
    \centering
    \begin{tabular}{|c|c|}
        \hline
        Gate & Trotter angle \\
        \hline
        $P_{1,2}$ & $\pi/4$ \\
        $\sqrt{P_{1,2}}$ & $\pi/8$ \\
        $(P_{1,2})^{1/4}$ & $\pi/16$ \\
        \hline
    \end{tabular}
    \caption{Exchange gates exactly realizable over
    Clifford+$T$ and their corresponding Trotter angles.}
    \label{tab:exact-swap-roots}
\end{table}

\subsection{Cartesian versus symmetry-adapted compilation}

At the level of ideal continuous gates, a standard implementation of the Heisenberg exchange interaction decomposes each isotropic bond Hamiltonian into its Cartesian components. 
\begin{equation}
\begin{split}
    e^{-i\delta t H_i}
    &= e^{-i\delta t
    (X_iX_{i+1}+Y_iY_{i+1}+Z_iZ_{i+1})} \\
    &= e^{-i\delta t X_iX_{i+1}}\,
       e^{-i\delta t Y_iY_{i+1}}\,
       e^{-i\delta t Z_iZ_{i+1}} .
\end{split}
\end{equation}
The choice of decomposition also affects the logical circuit structure. A naive implementation of the Cartesian decomposition uses six CNOT gates per nearest-neighbor bond, with two CNOTs for each Pauli-string rotation, whereas the fractional-SWAP construction described above uses four CNOT gates. These counts refer to the displayed constructions rather than to optimal lower bounds. More importantly for fault-tolerant compilation, the three continuous rotations in the Cartesian decomposition must be synthesized independently over Clifford+\(T\). Their approximation errors need not retain the cancellation responsible for the \(SU(2)\) symmetry of the ideal product. A symmetry-adapted compilation avoids this problem by ensuring that the compiled two-site operation remains within the \(SU(2)\)-symmetric operator algebra.

\section{Fault-Tolerant Compilation over Clifford+T}\label{sec:fault-tolerant}

In a fault-tolerant quantum computing architecture, one generally does not have access to arbitrary, high-fidelity continuous single-qubit rotations. Instead, these continuous operations must be compiled into a discrete, universal gate set, most commonly the Clifford+$T$ gate set. According to the Solovay--Kitaev theorem, because the Clifford+$T$ group is dense in $SU(2)$, any single-qubit unitary can be approximated to arbitrary precision $\epsilon$ using a sequence of these discrete gates~\cite{DawsonNielsen2006}. Modern synthesis algorithms significantly improve upon the asymptotic bounds of the original constructive proof. In particular, the \textit{gridsynth} algorithm implements the Ross--Selinger method for optimal or near-optimal ancilla-free approximation of single-qubit $R_z$ rotations~\cite{RossSelinger2016}; for brevity, we refer to the resulting circuit as an \textbf{approximant}. Exactly representable single-qubit Clifford+$T$ circuits can be placed in Matsumoto--Amano normal form~\cite{matsumoto2008representationquantumcircuitsclifford}.

However, standard synthesis algorithms are inherently symmetry-agnostic. Generic multiqubit methods can approximate a target by rounding it to a unitary over the Clifford+$T$ ring and then applying exact synthesis~\cite{Kliuchnikov2013UnitarySynthesis}, but such entrywise approximation does not guarantee preservation of commutation relations with the physical symmetry generators. Approximating an arbitrary continuous rotation through generic Clifford+T synthesis can therefore introduce truncation errors that explicitly break the continuous $SU(2)$ symmetries we wish to preserve. To circumvent this, one can attempt to implement specific Trotter steps \textbf{exactly} using Clifford+T circuits. For example, the exchange gates $\swap$, $\swap^{1/2}$, and $\swap ^{1/4}$ can all be compiled exactly without any synthesis errors. Notably, the smallest Trotter step exactly implementable in this decomposition corresponds to an angle of $\pi/16$, see Table \ref{tab:exact-swap-roots} and Appendix \ref{sec: swapexact}.  This provides an algorithmic advantage over a Cartesian decomposition, where the smallest Trotter step size is $\delta t = \pi/4$. Beyond these specific roots, however, the target unitary matrices no longer have entries that lie within the Clifford+T ring $\mathbb{Z}[1/\sqrt{2}, i]$, whose connection to exact single- and multiqubit Clifford+$T$ synthesis is characterized in Refs.~\cite{Kliuchnikov2013Exact,GilesSelinger2013}.

Because we are limited in the Trotter step sizes we can achieve with exact, symmetry-preserving gates, we must ask a broader question: Do there exist arbitrary discrete gates over the Clifford+T ring whose composition yields a unitary $U'$ that is arbitrarily close to our desired continuous rotation $U$ (i.e., $\|U - U'\| < \epsilon$), while strictly preserving the symmetry $[U', S] = 0$? 

Since our circuit is fundamentally composed of two-qubit interactions, we require that each approximate two-qubit unitary $U'^{(2)}$ strictly commutes with the local two-qubit symmetry generators. For the Heisenberg model, this is the total spin operator of the pair, $J^{(2)} = \vec{S}_1 + \vec{S}_2$. We demand:
\begin{align}
[U'^{(2)},J_\alpha^{(2)}]=0,\qquad \alpha=x,y,z.
\end{align}
This commutation relation implies that $U'^{(2)}$ and $({J^{(2)}}^2,J_z^{(2)})$ must share a simultaneous eigenbasis. For two spin-$1/2$ particles, this is naturally the singlet-triplet basis. Singlet-triplet measurements have also been proposed as the protected primitive in an $SU(2)$-invariant model of quantum computation~\cite{Freedman2021SymmetryProtected}. Here, in contrast, we use the same decomposition to characterize and compile symmetry-preserving time evolution. We define a basis transformation matrix $B$ that maps the computational basis to this symmetric basis
\begin{align}
    B\ket{00} &= \ket{T_1} = \ket{00} \\
    B\ket{01} &= \ket{T_{-1}} = \ket{11} \\
    B\ket{10} &= \ket{T_0} = \frac{1}{\sqrt{2}}(\ket{01}+\ket{10})\\
    B\ket{11} &= \ket{S} = \frac{1}{\sqrt{2}}(\ket{01}-\ket{10}) 
\end{align}
where $\ket{S}$ is the singlet state, and $\ket{T_i}$ denote the triplet states. 
In matrix form, $B$ is represented as:
\begin{equation}
   B = \begin{bmatrix}
 1 & 0 & 0 & 0\\
 0 & 0 & \frac{1}{\sqrt{2}} & \frac{1}{\sqrt{2}}\\
 0 & 0 & \frac{1}{\sqrt{2}} & -\frac{1}{\sqrt{2}}\\
 0 & 1 & 0 & 0
\end{bmatrix}
\end{equation}
Since $B$ is already in the matrix ring, this can be easily prepared using Clifford+T gates. In particular, using standard circuit identities, we can show that $B$ is
\begin{align}
    B = \swapp{1}{2} \cnott{2}{1}\text{\footnotesize CH}_{1,2}
\end{align}
where $\text{\footnotesize CH}$ is controlled Hadamard. 

By transforming the exact target unitary $U^{(2)}=e^{-i(XX+YY+ZZ)\delta t}$ into this basis, we obtain
\begin{equation}
    U^{(2)} = B D B^{\dagger}
\end{equation}
Because the isotropic Heisenberg Hamiltonian commutes with total spin, it is entirely diagonal in the singlet-triplet basis, not just block-diagonal. In particular, for this case, $D$ is given as
\begin{align}
    D=e^{-i\delta t}\cphase(4\delta t)
\end{align}

We now seek an approximate unitary $U'^{(2)}$ whose matrix elements belong entirely to the Clifford+T ring, such that it can be expressed as
\begin{equation}
    U'^{(2)} = B D' B^{\dagger}
\end{equation}
Here, $D'$ also has a specific form. To conserve $J^2$, the spin-0 and spin-1 sectors do not mix, which means that $D'$ is at least block diagonal. Since it also commutes with the spin observables $J_x,J_y,J_z$, it turns out that $D'$ must be diagonal, with the spin-1 sector having a common eigenvalue. This is because the spin-1 sector must commute with $J_x$ and $J_y$, which are not diagonal in the triplet basis. Thus $D'$ has the form:
\begin{align}
  D'=\diag(t,t,t,s) =t\diag(1,1,1,s/t) \\=\diag(1,1,1,e^{i\phi}) 
\end{align}
To satisfy the requirement that $U'{(2)}$ can be synthesized exactly over the Clifford+$T$ gate set, the matrix entries evaluated in the computational basis must yield values restricted to the ring $\mathbb{Z}[1/\sqrt{2}, i]$. Since $B$ is already in the ring, this imposes the following constraint on $D$.
\begin{equation}
    \frac{a+b\sqrt{2}}{\sqrt{ 2^k}} + i \frac{c+d\sqrt{2}}{\sqrt{ 2^k}} = e^{i \phi}
\end{equation}
with $a,b,c,d \in \mathbb{Z}$. As we prove in Appendix \ref{sec:nogo}, this imposes a severe number-theoretic restriction. It is mathematically impossible to find solutions for arbitrary angles $\phi$, effectively restricting this diagonal, symmetry-preserving Clifford+$T$ synthesis to a sparse, finite set of discrete angles. These include the angles $\phi = \pi,\pi/2, \pi/4$, which, as shown previously, can be obtained exactly. This obstruction depends on the fault-tolerant gate set: adjoining higher cyclotomic phase gates changes the ring of exactly synthesizable multiqubit matrices~\cite{Amy2024Cyclotomic}. Such Clifford-cyclotomic extensions may therefore admit additional exact symmetry-preserving angles, although they still do not provide a continuously parameterized family of exact symmetry preserving rotations.

\section{Adding Measurements: The LCU Method}
\label{sec:lcu}

As established, unitary Clifford+T synthesis is insufficiently dense to generate symmetry-preserving Trotter steps for arbitrary, infinitesimal time increments. Adding mid-circuit measurements allow us to overcome this constraint and modern fault-tolerant quantum architectures allow mid-circuit measurements and classical feed-forward. By incorporating these features, we can perform conditional, probabilistic operations using ancilla qubits. Our construction builds on on the linear-combination-of-unitaries framework~\cite{Berry2015LCU} and repeat-until-success synthesis~\cite{PaetznickSvore2014,Bocharov2015RUS}.
We can perform a $\cphase$ gate either by composing three $R_z$ gates or by synthesizing the $\cphase$ gate directly. 
The non symmetry-preserving version, which we call the \textbf{default} scheme, uses $\textit{gridsynth}$\cite{RossSelinger2016} to generate an $R_z$ gate, and we use three of the same $R_z$ gates to produce a $\cphase$. The symmetry preserving versions require an ancilla.  For instance, we can use prior methods ~\cite{Bocharov2015RUS} to generate an axial $R_z$ gate, and compile to get a $\cphase$, which we refer to as \textbf{BRS 2015} scheme. We can also directly synthesize a $\cphase$ gate using ancilla and conditional logic, and we call this method the \textbf{LCU} method. We will develop the \textbf{LCU} in detail, and also apply both known methods i.e. \textbf{default} and \textbf{BRS 2015} and analyze their effect on symmetry.  

For the convention
\begin{align}
    \cphase(\theta)
    &=\operatorname{diag}(1,1,1,e^{i\theta}) \\
    &=e^{i\theta/2}e^{-i(\theta/2)CZ},
    \label{eq:cphase-cz-identity}
\end{align}
the global phase $e^{i\theta/2}$ can be ignored, so the LCU construction only needs to approximate $e^{-i(\theta/2)CZ}$. For a Heisenberg Trotter step of duration $\delta t$, the target angle in Eq.~\eqref{eq:cphase-cz-identity} is $\theta=4\delta t$.

\begin{itemize}
    \item $\prep$: Initialize the ancilla in $\ket{0}_a$ and apply
    \begin{align}
        \prep
        &=2^{-l/2}
        \begin{bmatrix}
            z & y \\
            -y^* & z^*
        \end{bmatrix},
        \\
        \prep\ket{0}_a
        &=2^{-l/2}
        \bigl(z\ket{0}_a-y^*\ket{1}_a\bigr).
    \end{align}
    Here $z,y\in\mathbb{Z}[\omega]$, with
    $\omega=e^{i\pi/4}$, and unitarity requires
    \begin{equation}
        \lvert z\rvert^2+\lvert y\rvert^2=2^l.
    \end{equation}

    \item $\select$: Conditioned on the ancilla being in
    $\ket{1}_a$, apply $-i\,\mathrm{CZ}$ to the two data qubits using an $i\mathrm{CCZ}$ gate:
    \begin{equation}
        \select
        =\ket{0}_a\!\bra{0}_a\otimes I
        +\ket{1}_a\!\bra{1}_a
        \otimes\bigl(-i\,\mathrm{CZ}\bigr).
    \end{equation}
    The resulting state is
    \begin{align}
        (\select\circ\prep)\ket{0}_a\ket{\psi}
        =2^{-l/2}\Bigl(
            z\ket{0}_a\ket{\psi}
            +\\iy^*\ket{1}_a\,\mathrm{CZ}\ket{\psi}
        \Bigr).
    \end{align}

    \item $\prep^\dagger$ and $\measure$: Uncomputing and measuring
    the ancilla yields the two unnormalized Kraus operators
    \begin{align}
        K_0
        &=2^{-l}\bigl(
            \lvert z\rvert^2 I
            -i\lvert y\rvert^2\mathrm{CZ}
        \bigr),
        \label{eq:lcu-success-kraus}\\
        K_1
        &=2^{-l}zy^*
        \bigl(I+i\,\mathrm{CZ}\bigr).
        \label{eq:lcu-failure-kraus}
    \end{align}

    For outcome $0$, the normalized operation is
    \begin{align}
        \frac{
            \lvert z\rvert^2I
            -i\lvert y\rvert^2\mathrm{CZ}
        }{
            \sqrt{\lvert z\rvert^4+\lvert y\rvert^4}
        }
        &=
        \cos\!\left(\frac{\theta'}{2}\right)I
        -i\sin\!\left(\frac{\theta'}{2}\right)\mathrm{CZ}
        \nonumber\\
        &=e^{-i(\theta'/2)\mathrm{CZ}},
        \label{eq:lcu-success-unitary}
    \end{align}
    where
    \begin{align}
        \cos\!\left(\frac{\theta'}{2}\right)
        &=
        \frac{\lvert z\rvert^2}
        {\sqrt{\lvert z\rvert^4+\lvert y\rvert^4}},
        \\
        \sin\!\left(\frac{\theta'}{2}\right)
        &=
        \frac{\lvert y\rvert^2}
        {\sqrt{\lvert z\rvert^4+\lvert y\rvert^4}}.
    \end{align}
    Thus, up to the global phase $e^{i\theta'/2}$, the successful
    branch implements $\cphase(\theta')$. We choose $z$ and $y$ to
    minimize the internal half-angle error as
\begin{equation}
    \epsilon_{\mathrm{half}}
    :=
    \left|\frac{\theta'-\theta}{2}\right|.
    \label{eq:lcu-angular-error}
\end{equation}
The reported LCU error is the full \(\cphase\)-angle error,
\begin{equation}
    \epsilon_{\mathrm{LCU}}
    :=
    |\theta'-\theta|
    =
    2\epsilon_{\mathrm{half}}.
\end{equation}
Further details are given in Appendix~\ref{section: implementation}.

    The success probability is independent of the data state:
    \begin{align}
        p_{\mathrm{succ}}
        &=
        \frac{\lvert z\rvert^4+\lvert y\rvert^4}{2^{2l}}
        \geq\frac{1}{2}.
        \label{eq:lcu-success-probability}
    \end{align}
    The bound follows from
    $\lvert z\rvert^2+\lvert y\rvert^2=2^l$.

    For outcome $1$, Eq.~\eqref{eq:lcu-failure-kraus} gives
    \begin{equation}
        \frac{I+i\,\mathrm{CZ}}{\sqrt{2}}
        =e^{i\pi\mathrm{CZ}/4}
        =e^{i\pi/4}\,\mathrm{CS}^{\dagger}.
    \end{equation}
    Applying $\mathrm{CS}$ corrects this branch, after which the
    protocol can be repeated. Both branches are diagonal functions
    of $\mathrm{CZ}$ and therefore act only along the $z$ axis, thus $BK_iB^{\dagger},\{i=0,1\}$ acts identically on all three triplet states and commutes with the total spin. 
\end{itemize}

It is important to distinguish exact circuit implementation from exact
implementation of the target angle. For a selected triple \((z,y,l)\), the
PREP, SELECT, correction, and uncomputation operations are implemented
exactly over Clifford+\(T\), assuming ideal logical gates and measurements.
After any unsuccessful outcome is corrected and the protocol is repeated,
the completed RUS procedure implements
\(\cphase(\theta')\) exactly. In general, however,
\(\theta'\neq\theta\), so the implemented gate approximates the desired
\(\cphase(\theta)\) with coherent angle error
\[
    \epsilon_{\mathrm{LCU}}=|\theta'-\theta|.
\]
Thus the compilation preserves the \(SU(2)\) symmetry exactly even though it
only approximates the desired evolution within the symmetry-preserving
operator algebra. The measurement outcome affects the number of attempts,
but not the angle implemented by the completed protocol.

\subsection{Search over cyclotomic integers}

To find a low-cost approximation, write
\begin{align}
    z=a\omega^3+b\omega^2+c\omega+d,
    \qquad \omega=\frac{1+i}{\sqrt{2}},
\end{align}
with $a,b,c,d\in\mathbb{Z}$. Its norm is
\begin{align}
    |z|^2
    &=a^2+b^2+c^2+d^2
      +\sqrt{2}\bigl[d(c-a)+b(a+c)\bigr]\\
    &=A+B\sqrt{2}
\end{align}
where $A=a^2+b^2+c^2+d^2$ and $B=d(c-a)+b(a+c)$. The finite search used in this work can be summarized as follows.

\begin{center}
\fbox{%
\begin{minipage}{0.92\columnwidth}
\textbf{Algorithm 1: Search for a PREP approximation}

\textbf{Input:} Target $\cphase$ angle $\theta$, error tolerance $\epsilon_{\max}$, denominator exponents $l\in\mathcal{L}$, and an integer coefficient bound $C$.

\begin{enumerate}
    \item Enumerate $z=a\omega^3+b\omega^2+c\omega+d$ for $a,b,c,d\in[-C,C]$ and store each element by its norm pair $(A_z,B_z)$, where $|z|^2=A_z+B_z\sqrt{2}$.
    \item For each $l\in\mathcal{L}$ and each stored $z$, solve the complementary norm equation
    \begin{align*}
        |y|^2=2^l-|z|^2
        =(2^l-A_z)-B_z\sqrt{2}
    \end{align*}
    by looking up the norm pair $(2^l-A_z,-B_z)$ in the same table.
    \item For every matched triple $(z,y,l)$, compute
    \begin{align*}
        \frac{\theta'}{2}
        =\arctan\!\left(\frac{|y|^2}{|z|^2}\right),
        \qquad
        \epsilon=|\frac{\theta-\theta'}{2}|,
    \end{align*}
    together with $p_{\mathrm{succ}}$ from Eq.~\eqref{eq:lcu-success-probability}.
    \item Retain candidates satisfying $\epsilon\leq\epsilon_{\max}$ and select the one with the highest $p_{\mathrm{succ}}$.
\end{enumerate}

\textbf{Output:} A triple $(z,y,l)$ defining $U_{\mathrm{PREP}}$ and the implemented angle $\theta'$.
\end{minipage}}
\end{center}

If $\operatorname{cde}(U_{\mathrm{PREP}})$ denotes the common denominator exponent of the PREP unitary after cancellation of common factors, the synthesis estimate used here is \cite{gosset_tcount}:
\begin{align}
    T_{\mathrm{PREP}}=2\operatorname{cde}(U_{\mathrm{PREP}}).
\end{align}
This is the cost of the PREP circuit only, not the total expected T-count of the complete RUS protocol.  We use T-count here to maintain a uniform comparison between compilation strategies.

\subsection{Expected T-count of the LCU protocol}

Let $\overline{T}_{\mathrm{LCU}}$ denote the expected T-count
of one successful $\cphase$ implementation. Every attempt applies
$\prep$, $\select$, and $\prep^\dagger$, at a cost
\begin{equation}
    T_{\mathrm{attempt}}
    =2T_{\mathrm{PREP}}+T_{\mathrm{SELECT}}.
\end{equation}
Because the number of attempts is geometrically distributed, the
expected numbers of attempts and failures are, respectively,
\begin{equation}
    \frac{1}{p_{\mathrm{succ}}}
    \qquad\text{and}\qquad
    \frac{1-p_{\mathrm{succ}}}{p_{\mathrm{succ}}}.
\end{equation}
The expected total cost is therefore
\begin{equation}
    \overline{T}_{\mathrm{LCU}}
    =
    \frac{
        2T_{\mathrm{PREP}}
        +T_{\mathrm{SELECT}}
        +(1-p_{\mathrm{succ}})T_{\mathrm{corr}}
    }{p_{\mathrm{succ}}}.
    \label{eq:expected-rus-t-count}
\end{equation}
Here $p_{\mathrm{succ}}$ is given by
Eq.~\eqref{eq:lcu-success-probability}.

The $\select$ operation is a CCZ followed by a Clifford
$S^\dagger$ gate on the ancilla. An exact ancilla-free CCZ---equivalently, a CCX
up to Hadamard gates---has T-count seven
\cite{gosset_tcount}. The
$\mathrm{CS}$ correction has T-count three. Hence
\begin{align}
    \overline{T}_{\mathrm{LCU}}
    &=
    \frac{
        2T_{\mathrm{PREP}}
        +7+3(1-p_{\mathrm{succ}})
    }{p_{\mathrm{succ}}}
    \nonumber\\
    &=
    \frac{
        4\operatorname{cde}(U_{\mathrm{PREP}})
        +7+3(1-p_{\mathrm{succ}})
    }{p_{\mathrm{succ}}},
    \label{eq:lcu-explicit-t-count}
\end{align}
where
$T_{\mathrm{PREP}}
 =2\operatorname{cde}(U_{\mathrm{PREP}})$.

Both $T_{\mathrm{PREP}}$ and $p_{\mathrm{succ}}$ depend on the
selected triple $(z,y,l)$. Increasing the allowed range of $l$
enlarges the candidate set and can improve the approximation
accuracy, but it does not necessarily increase the T-count
monotonically. The cost is determined jointly by the reduced common
denominator exponent and the success probability.

Before presenting the numerical benchmarks, we summarize the error
conventions used for the three compilation methods in
Table~\ref{tab:error-conventions}. These quantities are not identical:
\(\epsilon_{\mathrm{GS}}\) is an operator-norm tolerance for each
single-qubit rotation, whereas \(\epsilon_{\mathrm{BRS}}\) and
\(\epsilon_{\mathrm{LCU}}\) are full controlled-phase angle errors.
\begin{table}[t]
\centering
\begin{tabular}{lll}
\hline
Method & Error parameter & Symmetry guarantee \\
\hline
Default GridSynth
&
$\|\widetilde R_z-R_z\|\leq\epsilon_{\mathrm{def}}$
&
None
\\
BRS 2015
&
$\epsilon_{\mathrm{BRS}}
 =|\theta'_{\mathrm{BRS}}-\theta|$
&
Exact
\\
LCU
&
$\epsilon_{\mathrm{LCU}}
 =|\theta'_{\mathrm{LCU}}-\theta|$
&
Exact
\\
\hline
\end{tabular}
\caption{Error conventions used for the three compilation methods.
The GridSynth tolerance applies to each synthesized single-qubit rotation;
the resulting controlled-phase error is bounded by
\(3\epsilon_{\mathrm{def}}\). For BRS 2015 and LCU,
the quoted error is the full controlled-phase angle error. }
\label{tab:error-conventions}
\end{table}

\begin{figure}
    \centering
    \includegraphics[width=\linewidth]{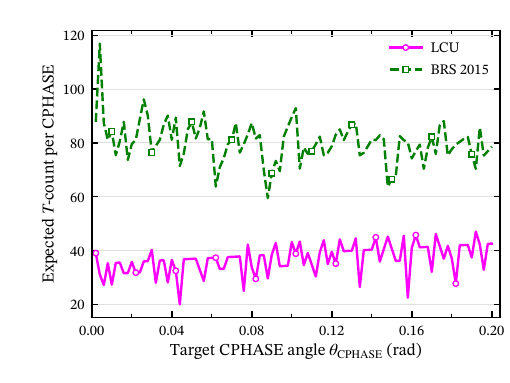}
    
    \caption{
Expected RUS \(T\)-count per \(\cphase\) gate for the BRS 2015 and
LCU methods at a common $\cphase$-angle tolerance,
\(\lvert\theta'_{\mathrm{CPHASE}}-\theta_{\mathrm{CPHASE}}\rvert
\leq 10^{-3}\).
At each target angle, we select the candidate with the minimum expected
\(T\)-count satisfying this tolerance. LCU has lower expected \(T\)-count
at every sampled angle. Over the displayed range, the mean expected
\(T\)-counts are \(36.9\) for LCU and \(80.3\) for BRS 2015, giving a
ratio of \(2.18\). For this plot and the rest of paper, we fix $\mathcal{L}$ and $C$ as 20.
}
    \label{fig:cphase-expected-t-count}
\end{figure}

\section{Numerical benchmarks on Heisenberg model}
\label{sec:numerical-benchmarks}
\subsection{Symmetry-preserving compilation of KPZ dynamics}
In classical systems, nonequilibrium physics is a source of rich phenomena and is often explained using concepts from hydrodynamics. One famous universality class that describes the dynamics in nonequilibrium systems is the Kardar--Parisi--Zhang universality class. It has been theoretically shown~\cite{Prosen_theory,DeNardis2020Solitons} that certain quantum systems with $SU(2)$ symmetry, such as the 1D Heisenberg model, possess features that belong to the KPZ universality class~\cite{Banerjee_2026,GoogleKPZ}. The stability of this superdiffusive behavior under weak integrability-breaking perturbations has also been investigated~\cite{DeNardis2021Stability}. Moreover, Trotterized versions of the model, which in general do not conserve energy, can also show KPZ scaling. Experimental studies have observed superdiffusive magnetization transport consistent with aspects of KPZ scalings~\cite{GoogleKPZ,Bloch_group_expt}.

\begin{figure}
    \centering
    \includegraphics[width=\linewidth]{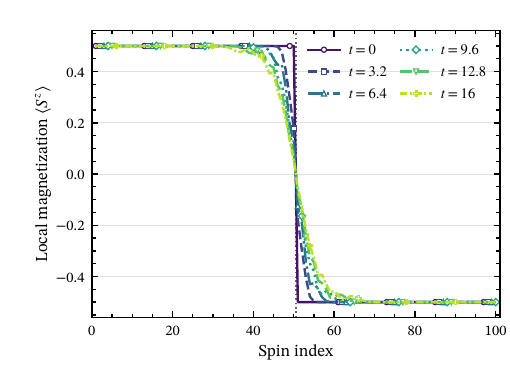}
    \caption{Pure, fully polarized domain-wall state obtained in the limit $\mu\rightarrow\infty$ at $t=0$, followed by its melting under symmetric Heisenberg evolution. Profiles are shown every 40 time steps from $t=0$ to $t=200\,\delta t$ with $\delta t=0.08$, corresponding to the physical snapshot times $t=0,3.2,6.4,9.6,12.8,$ and $16$.}
    \label{fig:domainwall}
\end{figure}

The Heisenberg model corresponds to the $\Delta = 1$ limit of the XXZ model
\begin{equation}
    H = \sum_i \left( X_i X_{i+1} + Y_i Y_{i+1} + \Delta Z_i Z_{i+1} \right)
\end{equation}
To study the dynamics of this model, the initial state can be chosen to be a polarized domain-wall ensemble~\cite{GoogleKPZ}:
\begin{equation}
    \rho(t=0) \propto (e^{2\mu S_z})^{\otimes N_Q/2} \otimes (e^{-2\mu S_z})^{\otimes N_Q/2}
\end{equation}
where $S_z=Z/2$, $N_Q$ is the total number of spins, and $\mu$ controls the polarization. For spins prepared thermally in a Zeeman field with splitting $h$, one may write $\mu=\beta h/2$, where $\beta=1/(k_B T)$. In the limit $\mu\rightarrow \infty$, this becomes a pure, fully polarized domain-wall state, with the two halves magnetized in opposite directions. We plot this state in Figure \ref{fig:domainwall}, and use it in our simulations.

Next, we evolve the system under the Heisenberg interaction. Under this evolution, we expect the domain wall to ``melt,'' resulting in transferred magnetization. At time $t=0$, the initial magnetization of the left half is $M_0$ (represented by the area under the curve in Fig.~\ref{fig:domainwall}); at time $t$, it decreases to $M(t)$. The transferred magnetization $\Delta M(t)$ is therefore
\begin{equation}
    \Delta M(t) = M_0 - M(t)
\end{equation}
Spin-current autocorrelation functions provide an alternative transport diagnostic and have been measured in digital simulations of XXZ chains across ballistic, superdiffusive, and diffusive regimes~\cite{Banerjee_2026}. Here we use transferred magnetization, which scales as
\begin{equation}
    \Delta M \propto t^{1/z}
\end{equation}
where
\begin{equation}
z = \begin{cases}
    1 & \text{if } \Delta < 1 \quad \text{(ballistic)} \\
    \tfrac{3}{2} & \text{if } \Delta = 1 \quad \text{(superdiffusive)} \\
    2 & \text{if } \Delta > 1 \quad \text{(diffusive)}
\end{cases}
\end{equation}
The regime in which we are interested is the Heisenberg limit, $\Delta=1$. 
We now probe how the transport exponent depends on synthesis errors introduced by a default Clifford+$T$ compiler. For a single bond, define
\begin{equation}
    K_j=X_jX_{j+1}+Y_jY_{j+1}+Z_jZ_{j+1},
\end{equation}
so that the ideal two-site gate is $U_j(\delta t)=e^{-i\delta t K_j}$. For any compiled gate $\widetilde U_j$ sufficiently close to $U_j$, its effective generator is
\begin{equation}
    \widetilde h_j=\frac{i}{\delta t}\log(\widetilde U_j),
    \label{eq:compiled-effective-generator}
\end{equation}

For the symmetry-preserving construction, the target $\cphase $ angle is $\theta=4\delta t$. The error in the $\cphase$ gate for LCU and BRS 2015 methods is respectively given as $\epsilon_{LCU}$ and $\epsilon_{BRS}$. Thus the evolution time for each is given as
\begin{align}
    \delta t ' = \frac{\theta' }{4} = \delta t +\frac{\Delta \theta}{4}
\end{align}
where for the LCU method
\begin{align}
    \Delta \theta= \epsilon_{LCU}
\end{align}
and for the BRS 2015 method
\begin{align}
    \Delta \theta= \epsilon_{BRS}
\end{align}

The implemented bond gate and its effective generator are consequently
\begin{align}
    \widetilde U_{j,\mathrm{sym}}
    &=e^{-i\delta t'K_j},\\
    \widetilde h_{j,\mathrm{sym}}
    &=\frac{\delta t'}{\delta t}K_j
      =\frac{\theta'}{\theta}K_j.
    \label{eq:lcu-rescaled-generator}
\end{align}
which for LCU is
\begin{align}
h_{j,\mathrm{LCU}}
    &
      =1+\frac{\epsilon_{LCU}}{4}K_j
\end{align}
and BRS 2015 is
\begin{align}
h_{j,\mathrm{BRS}}
    &
      =1+\frac{\epsilon_{BRS}}{4}K_j
\end{align}

Thus the angle error changes the exchange strength but introduces no new Pauli operators. In particular, $[\widetilde h_{j,\mathrm{sym}},J_\alpha]=0$ for $\alpha=x,y,z$. When the same angle is used on every bond and at every Trotter step, this is a uniform rescaling of the simulated time.   

The precision parameter of the default \textit{gridsynth} compiler has a different interpretation. If $\widetilde R_z(\alpha)$ is the Clifford+$T$ approximation to $R_z(\alpha)$, the input tolerance $\epsilon_{\mathrm{def}}$ guarantees~\cite{RossSelinger2016}
\begin{equation}
    \left\|\widetilde R_z(\alpha)-R_z(\alpha)\right\|
    \leq\epsilon_{\mathrm{def}}.
    \label{eq:gridsynth-rz-bound}
\end{equation}

It does not specify either the direction or the magnitude of the off-axis component of the particular approximant returned by the compiler. We can upper bound the gate error for Trotter steps generated via gridsynth using a telescoping sum (see Appendix \ref{section: implementation} ) as $\epsilon_{\mathrm{def}} \leq 3 \epsilon$, where $\epsilon_{\mathrm{def}}$ is the two qubit gate error for the default method.  

However, note that the magnitude of the symmetry-breaking component need not be a monotonic function of the requested $\epsilon$.

To illustrate the resulting perturbation, we consider the parameters used in the low-precision transport comparison, $\delta t=0.08$ and $\epsilon_{\mathrm{def}}=2\times10^{-2}$. For the resulting GridSynth approximant,
\begin{align}
    \left\|\widetilde U_{j,\mathrm{def}}-U_j\right\|
    &=1.48\times10^{-2}<3\epsilon_{\mathrm{def}}.
\end{align}
Using Eq.~\eqref{eq:compiled-effective-generator} and removing the identity contribution associated with a global phase, the effective two-site generator is
\begin{align}
    \widetilde h_{j,\mathrm{def}}
    \simeq{}&1.05\,K_j
    +0.043\,Z_jY_{j+1}\nonumber\\
    &-0.043\,Y_jZ_{j+1}
    -0.043\,Y_jX_{j+1}\nonumber\\
    &+0.033\,Z_j
    +0.0033\,Z_{j+1}\nonumber\\
    &+0.0022\,X_jY_{j+1}.
    \label{eq:gridsynth-example-generator}
\end{align}
Terms with coefficient magnitude below $1.5\times10^{-4}$ have been omitted from Eq.~\eqref{eq:gridsynth-example-generator}. The first term is an $SU(2)$-preserving change of the exchange strength. The remaining terms are symmetry breaking; for example,
\begin{equation}
    [Z_jY_{j+1},J_z]=iZ_jX_{j+1}\neq0.
\end{equation}

The matrix-product calculations implement these effective two-site generators using a fixed Trotter schedule. For $r=t/\delta t$ steps, the two evolutions take the form
\begin{align}
    \widetilde U_{\mathrm{sym}}(t)
    &=\left[\prod_j e^{-i\delta t'K_j}\right]^r,\\
    \widetilde U_{\mathrm{def}}(t)
    &=\left[\prod_j e^{-i\delta t\widetilde h_{j,\mathrm{def}}}\right]^r,
\end{align}
where the product follows the same bond ordering in both simulations. We then extract the transport exponent from the transferred magnetization $\Delta M(t)$.
\subsubsection{Transport accuracy and fault-tolerant cost}
\begin{figure*}[t]
    \centering
    \begin{minipage}{0.48\textwidth}
        \centering
        \includegraphics[width=\linewidth]{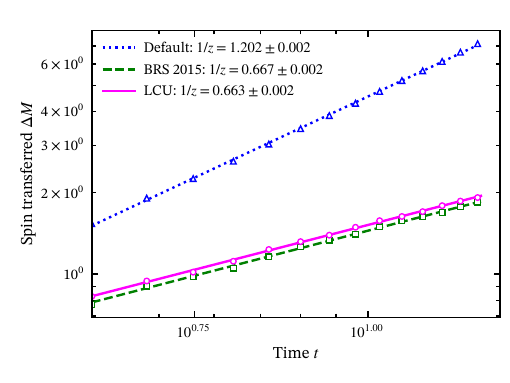}
        \par\smallskip
        \textbf{(a)}
    \end{minipage}\hfill
    \begin{minipage}{0.48\textwidth}
        \centering
        \includegraphics[width=\linewidth]{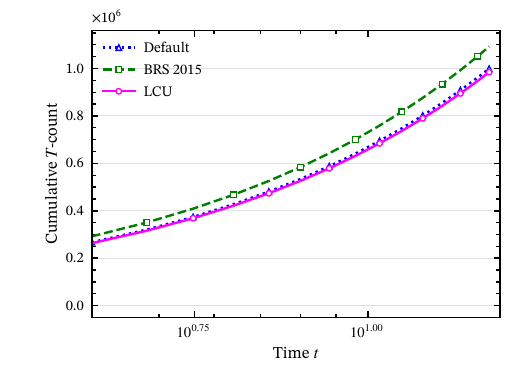}
        \par\smallskip
        \textbf{(b)}
    \end{minipage}
    \caption{Comparable $T$-count, different transport exponents. (a) We choose synthesis precisions $\epsilon=2\times10^{-2}$ for both the default and BRS 2015 methods, while the LCU cost uses $\epsilon_{\mathrm{LCU}}=4\times10^{-5}$. At these values, the $T$-count is nearly the same, but the exponent is different for the symmetry breaking and symmetry preserving methods: $1.202\pm 0.002$ for the default method, $0.667\pm0.002$ for BRS 2015, and $0.663\pm0.002$ for LCU, which is extracted as power law fit from the log-log plots (b) Synthesis precisions are chosen so that all three methods have similar T counts. The simulations use $N=100$, $\delta t=0.08$, and total evolution time $T=15$. We plot time on a log scale. }
    \label{fig:kpz-accuracy-cost_panel1}
\end{figure*}

We compare the three compilation strategies in two regimes. First, we fix the synthesis precision to $\epsilon=2\times10^{-2}$ for the BRS 2015
and default method, and increase the precision of LCU to $4\times 10^{-5}$ in the transport comparison of Fig.~\ref{fig:kpz-accuracy-cost_panel1}(a). We choose these precision values in order to keep T count same for all three methods, and from Figure \ref{fig:kpz-accuracy-cost_panel1}
, it can be seen that symmetry-preserving methods give a much better estimate of the KPZ exponent. In particular,  $0.667\pm0.002$, and $0.663\pm0.002$ for the BRS 2015 and LCU methods, respectively, while for default, it is $1/z=1.202\pm0.002$. 

\begin{figure*}
    \begin{minipage}{0.48\textwidth}
        \centering
        \includegraphics[width=\linewidth]{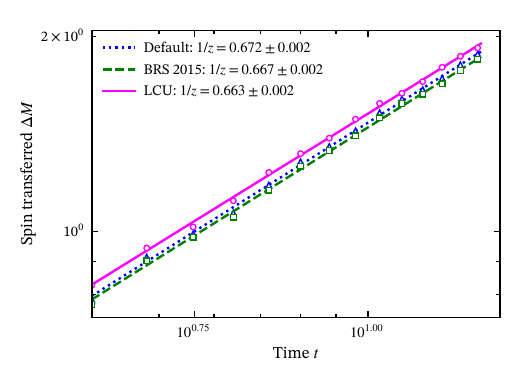}
        \par\smallskip
        \textbf{(a)}
    \end{minipage}\hfill
    \begin{minipage}{0.48\textwidth}
        \centering
        \includegraphics[width=\linewidth]{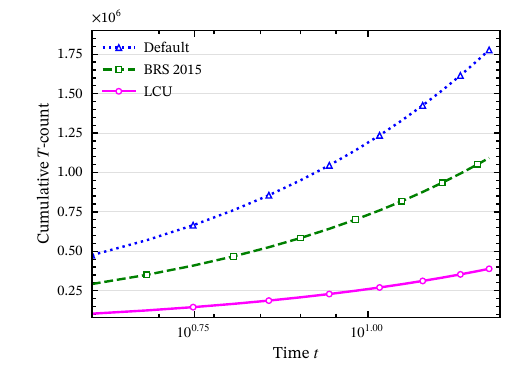}
        \par\smallskip
        \textbf{(b)}
    \end{minipage}
    \caption{Similar exponents, different $T$-count (a) Independent matched-accuracy simulations give $1/z=0.672\pm0.002$ for the default method at $\epsilon_{\mathrm{def}}=2\times10^{-3}$, $0.667\pm0.002$ for BRS 2015 at $\epsilon_{\mathrm{BRS}}=2\times10^{-2}$, and $0.663\pm0.002$ for LCU at requested $\epsilon_{\mathrm{LCU}}=2\times10^{-2}$, where the exponent is plotted using a power law fit on the log-log plot. (b) At these precisions, BRS 2015 has roughly $60$ percent $T$ count compared to default, and LCU has $20$ percent $T$ count compared to the default method. Note that these numbers are obtained for the desired $\delta t$, and $T$ cost may depend on the chosen step size.  The simulations use $N=100$, $\delta t=0.08$, and total evolution time $T=15$. We plot time on a log scale. }
    \label{fig:kpz-accuracy-cost}
\end{figure*}

Second, increasing the precision of the default method to $\epsilon_{\mathrm{def}}=2\times10^{-3}$ restores $1/z\simeq2/3$ in Fig.~\ref{fig:kpz-accuracy-cost}(a): independent simulations give $1/z=0.672\pm0.002$, $0.667\pm0.002$, and $0.663\pm0.002$ for the default, BRS 2015, and LCU methods, respectively.  The default calculation uses $\epsilon=2\times10^{-3}$, while the BRS 2015 and LCU calculation uses $\epsilon_{\mathrm{BRS}}=2\times10^{-2}$.  The cost comparison in Fig.~\ref{fig:kpz-accuracy-cost}(b) uses the same precisions, with a logarithmic time axis and a linear T-count axis.

As discussed above, when the same compiled angle is used on every bond and at every Trotter step, the synthesis error in either symmetry-preserving method produces a uniform rescaling of the exchange coupling by
$c=\theta'/\theta$. Neglecting product-formula error, the compiled dynamics therefore satisfies
$\Delta M_{\mathrm{comp}}(t)=\Delta M(ct)$. Consequently, an asymptotic power law
$\Delta M(t)\sim A t^\alpha$ becomes
\begin{align}
    \Delta M_{\mathrm{comp}}(t)
    \sim A(ct)^\alpha
    = A c^\alpha t^\alpha ,
\end{align}
so the rescaling changes the prefactor but not the asymptotic exponent. Thus, within this setting, the asymptotic exponent is independent of the synthesis precision, provided that the compiled angle remains uniform across bonds and steps.

The exponent is nevertheless extracted numerically over a finite time window, in which corrections to asymptotic scaling remain present. For example,
\begin{align}
    \Delta M(t)
    = A t^{2/3}\left(1+b t^{-\omega}+\cdots\right)
\end{align}
becomes
\begin{align}
    \Delta M_{\mathrm{comp}}(t)
    = A c^{2/3} t^{2/3}
      \left(1+b c^{-\omega}t^{-\omega}+\cdots\right).
\end{align}
Different values of $c$ therefore sample different effective-time windows and can yield slightly different finite-window fitted exponents. In addition, the rescaled interaction angle corresponds to an effective Trotter step $\delta t'=c\,\delta t$. Because BRS 2015 and LCU generally produce different values of $c$, they also incur slightly different finite-step Trotter corrections. These finite-time and Trotter effects account for the small difference between the exponents extracted for BRS 2015 and LCU, without implying different asymptotic transport exponents.

\begin{figure}[t]
    \centering
    \includegraphics[width=\linewidth]{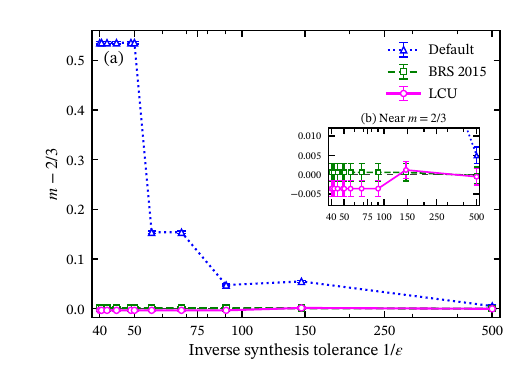}
    \caption{Deviation of the fitted transport slope $m=1/z$ from $2/3$ as a function of inverse synthesis tolerance. At lower $1/\epsilon$ (lower precision), the default method deviates significantly from $m$, while both symmetry preserving methods have much smaller errors. At higher $1/\epsilon$, symmetry preserving and violating methods converge. }
    \label{fig:kpz-inverse-epsilon}
\end{figure}

The dependence on synthesis precision is summarized more broadly in Fig.~\ref{fig:kpz-inverse-epsilon}. The fitted slope is denoted by $m=1/z$, so that $m-2/3$ measures its deviation from the expected Heisenberg value. The symmetry-preserving constructions remain close to $m=2/3$ across the precision range, while the default method approaches this value only as its synthesis precision is increased.
\subsubsection{\texorpdfstring{$m-2/3$ versus $T$-count}{m-2/3 versus T-count}}
\begin{figure}[t]
    \centering
    \includegraphics[width=\linewidth]{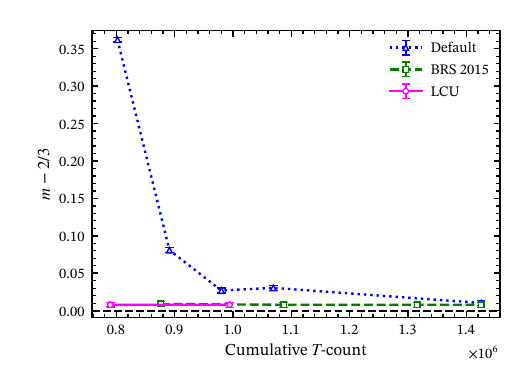}
    \caption{Deviation of the fitted transport slope from the KPZ value versus cumulative T-count. LCU method uses the range $10^{-6}\leq\epsilon_{\mathrm{LCU}}\leq10^{-4}$, while $\epsilon_{\mathrm{def}}$ and $\epsilon_{\mathrm{BRS}}$ ranges are given in the main text.}  
    
    \label{fig:kpz_t_count}
    
\end{figure}

The synthesis tolerance can be eliminated parametrically to display the
transport accuracy directly against the fault-tolerant resource cost.  For
each requested tolerance $\epsilon$, we pair the fitted deviation
$m(\epsilon)-2/3$ with
\begin{equation}
    T_{\mathrm{cum}}(\epsilon)
    =T_{\mathrm{bond}}(\epsilon)(N-1)\frac{t_{tot}}{\delta t},
    \label{eq:kpz-cumulative-t-count}
\end{equation}
where $T_{\mathrm{bond}}$ is the T-count of one compiled two-site update.  All
three datasets in Fig.~\ref{fig:kpz_t_count} use the same physical
parameters, $N=100$, $\delta t=0.08$, and total evolution time
$t_{tot}=12$, corresponding to 150 Trotter steps. 

Because the synthesis spectra of the three constructions are discrete and
have different useful tolerance windows, we use method-specific ranges.  The
default method uses
$2\times10^{-3}\leq\epsilon_{\mathrm{def}}\leq2.5\times10^{-2}$, the BRS
construction uses
$10^{-4}\leq\epsilon_{\mathrm{BRS}}\leq10^{-1}$, and the LCU construction is
searched over
$10^{-6}\leq\epsilon_{\mathrm{LCU}}\leq10^{-4}$.  Multiple
requested tolerances often select exactly the same Clifford+$T$ circuit, so
identical circuit points are consolidated rather than plotted repeatedly. At low T counts, either symmetry-preserving method gives a more accurate estimate of the critical exponent than using a non symmetry preserving method.

The tolerance ranges explored in
Figs.~\ref{fig:kpz-inverse-epsilon} and
\ref{fig:kpz_t_count} are subject to both synthesis
and numerical limitations. To resolve a nontrivial rotation,
the synthesis tolerance should be small compared with the
target rotation scale. Once the tolerance exceeds the distance
between the target gate and the identity, the identity becomes
an admissible approximation with zero $T$-count. This places a
practical upper limit on the useful tolerance range, with the
precise threshold depending on the gate and error convention.

The accessible range may also be limited by the bond dimension
required for accurate matrix-product-state simulations. In the
default compilation, looser tolerances permit larger
symmetry-breaking perturbations, which can alter entanglement
growth and increase the computational cost. In the
symmetry-preserving constructions, the angle error instead
rescales the exchange coupling by $\kappa=\theta'/\theta$.
When $\kappa>1$, a fixed nominal evolution time corresponds
to a longer effective evolution, which can likewise increase
the required bond dimension. Neither effect is necessarily
monotonic in the requested tolerance. The observed trend
suggests that the default method may exhibit larger deviations
from the KPZ exponent at looser tolerances, but this
verification remains limited by computational capacity. 
\subsection{Global fidelity and symmetry sector fidelity}

We first compare the three compilation strategies using state fidelity and
fault-tolerant resource cost.  The methods are (i) \textbf{default}, symmetry-agnostic
single-qubit synthesis, (ii) the \textbf{BRS 2015} \cite{Bocharov2015RUS} construction, and (iii) our
symmetry-preserving LCU construction.  To avoid choosing an initial state that
is accidentally insensitive to the synthesis error, we use a fixed Haar-random
state for the global-fidelity comparisons and use the same state,
system size, and total evolution time for all three methods.  We distinguish
the total infidelity
\begin{equation}
    1-F_{\mathrm{tot}}^{(m)}
    =1-\left|\braket{\psi_{\mathrm{exact}}(T)|
    \widetilde{\psi}_{m}(T)}\right|^2,
    \label{eq:total-infidelity-comparison}
\end{equation}
from the synthesis infidelity
\begin{equation}
    1-F_{\mathrm{syn}}^{(m)}
    =1-\left|\braket{\psi_{\mathrm{Trot}}(T)|
    \widetilde{\psi}_{m}(T)}\right|^2.
    \label{eq:synthesis-infidelity-comparison}
\end{equation}
Here, $\ket{\psi_{\mathrm{exact}}(T)}$ is obtained from exact Heisenberg
evolution, $\ket{\psi_{\mathrm{Trot}}(T)}$ is obtained from the product formula
with exact continuous gates, and $\ket{\widetilde{\psi}_{m}(T)}$ is the state
produced by compilation method $m$.  The exact-gate Trotter curve therefore
isolates the product-formula error, while Eq.~\eqref{eq:synthesis-infidelity-comparison}
isolates the change caused by gate synthesis.  These two infidelities are not,
in general, additive because Trotter and synthesis errors can interfere
coherently.

\begin{figure*}[t]
    \centering
    \includegraphics[width=0.48\textwidth]{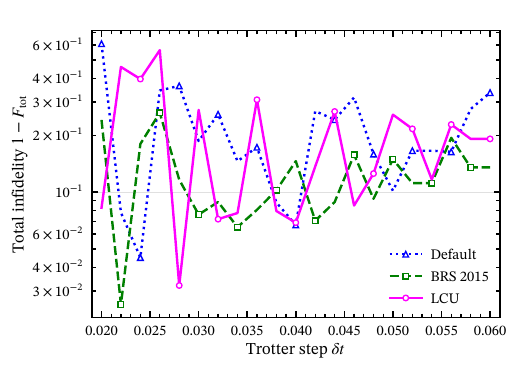}
    \hfill
    \includegraphics[width=0.48\textwidth]{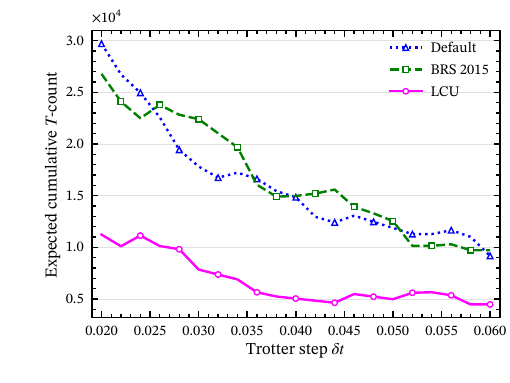}
    \caption{Constant-bias compilation for a fixed Haar-random initial state.
    \textbf{(a)} Total infidelity relative to exact Heisenberg evolution.
    \textbf{(b)} Expected total $T$-count for the same circuits.  The
    simulations use $N=10$, total time $t_{tot}=1$, synthesis precision $\epsilon=10^{-2}$ for all three methods.
    At the larger step sizes the three methods have comparable total
    infidelity, while the LCU construction retains a substantially smaller
    expected $T$-count.}
    \label{fig:constant-random-fidelity-cost}
\end{figure*}

We begin with a constant synthesis bias, for which the signed error in the
implemented angle has the same sign at every Trotter step. In the regime where
the synthesis error is smaller than or comparable to the Trotter error, the
three methods approach the same total-infidelity curve as the physical Trotter
step is increased (see Figure \ref{fig:constant-random-fidelity-cost}). Despite their comparable accuracy in
the Trotter-dominated regime, the methods require substantially different numbers of
$T$ gates.  Thus, even when ordinary state fidelity does not favor one
construction, fault-tolerant resource accounting does: our LCU construction
reaches the same accuracy at a lower expected $T$-count. 

\begin{figure*}[t]
    \centering
    \includegraphics[width=0.48\textwidth]{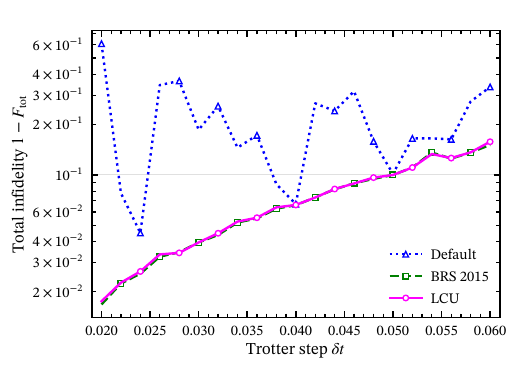}
    \hfill
    \includegraphics[width=0.48\textwidth]{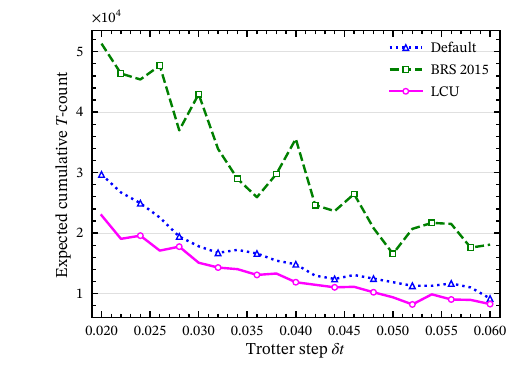}
    \caption{Alternating-bias compilation for the same fixed Haar-random initial
    state used in Fig.~\ref{fig:constant-random-fidelity-cost}.  \textbf{(a)} Total infidelity relative
    to exact Heisenberg evolution.  Alternating over- and under-rotations make
    the BRS 2015 and LCU errors nearly coincide.  \textbf{(b)} Expected total
    $T$-count, showing that the BRS 2015 construction requires the most
    resources.  The simulations use $N=10$, $t_{tot}=1$, $\epsilon=10^{-2}$ for all three methods. }
    \label{fig:alternating-singlet-fidelity-cost}
\end{figure*}

We also consider an
alternating schedule in which successive steps use over and under-rotations.
The leading signed-angle errors then cancel pairwise, strongly suppressing
their coherent accumulation.  For the same fixed Haar-random initial state used
in Fig.~\ref{fig:constant-random-fidelity-cost}, the alternating BRS 2015 and
LCU curves are nearly indistinguishable and lie below the default
symmetry-breaking synthesis curve.  This cancellation is not
free, however. In searching for  a suitable pair of approximants on opposite sides of the target
angle, the T-count for approximants that correspond to angles $\theta +\epsilon$ and $\theta - \epsilon$ is higher, thus comparing with Figure \ref{fig:constant-random-fidelity-cost}, the T-count is higher for both BRS 2015 and LCU in Figure \ref{fig:alternating-singlet-fidelity-cost}. However, LCU can still achieve a better fidelity than the default method, while having the same T-count.  

Global fidelity alone does not determine whether the compiled dynamics obeys
the symmetry of the target Hamiltonian.  To expose this distinction, we next
initialize four spins in a product of two singlets,
\begin{equation}
    \ket{\psi_0}=\ket{S}_{01}\otimes\ket{S}_{23},
\end{equation}
which lies in the total-spin $S=0$ sector.  Let $P_{S=0}$ denote the projector
onto this sector.  For four spin-$1/2$ particles, the allowed eigenvalues of
$J^2$ are $0$, $2$, and $6$, and the projector can be written as
\begin{equation}
    P_{S=0}=\frac{(J^2-2I)(J^2-6I)}{12}.
    \label{eq:singlet-sector-projector}
\end{equation}
We quantify leakage from the physical sector by
\begin{equation}
    \mathcal{L}^{(m)}(T)
    =1-\bra{\widetilde{\psi}_{m}(T)}P_{S=0}
    \ket{\widetilde{\psi}_{m}(T)}.
    \label{eq:singlet-sector-leakage}
\end{equation}
where $\ket{\tilde{\psi}_m(T)}$ is the Trotter-evolved state under method $m$ (BRS, default, LCU) at time T. 
This is a stricter diagnostic than projection onto the
$S^z_{\mathrm{tot}}=0$ subspace: conservation of total magnetization tests only
the $U(1)$ subgroup, whereas Eq.~\eqref{eq:singlet-sector-leakage} detects
transitions between different total-spin irreducible sectors.  The leakage
comparison shows the physical information missed by global fidelity.  Two
compiled states may have similar overlap with the target state while differing in the probability that they remain in the symmetry sector selected by
the initial condition.

\begin{figure}[t]
    \centering
    \includegraphics[width=\linewidth]{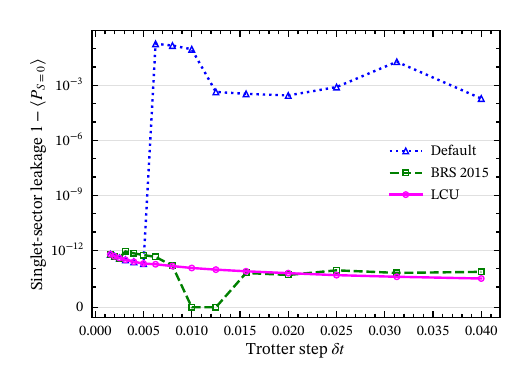}
    \caption{Leakage from the total-spin singlet sector for the initial state
    $\ket{S}_{01}\otimes\ket{S}_{23}$ at $N=4$, $T=1$, and
    $\epsilon=10^{-2}$.  The default symmetry-agnostic synthesis populates
    sectors with $S>0$, whereas the BRS 2015 and LCU constructions preserve
    the $S=0$ sector up to numerical precision. The vertical axis uses a symmetric logarithmic scale, with a linear
region near zero. }
    \label{fig:total-spin-sector-leakage}
\end{figure}

Finally, we separate symmetry leakage from distortion of the state within the
physical sector.  We use synthesis tolerances
$\epsilon_{\mathrm{LCU}}=2\times10^{-5}$ and
$\epsilon_{\mathrm{BRS}}=2\times10^{-3}$ for the LCU and BRS
constructions, respectively, and compare them using
the conditional in-sector fidelity

\begin{align}
    p_{\mathrm{ref}}^{(0)}
    &=
    \bra{\psi_{\mathrm{ref}}(T)}
    P_{S=0}
    \ket{\psi_{\mathrm{ref}}(T)},\\
    p_m^{(0)}
    &=
    \bra{\widetilde{\psi}_m(T)}
    P_{S=0}
    \ket{\widetilde{\psi}_m(T)},
\end{align}
so that
\begin{equation}
    F_{\mathrm{cond}}^{(m)}
    =
    \frac{
        \left|
        \bra{\psi_{\mathrm{ref}}(T)}
        P_{S=0}
        \ket{\widetilde{\psi}_m(T)}
        \right|^2
    }{
        p_{\mathrm{ref}}^{(0)}p_m^{(0)}
    }.
    \label{eq:conditional-singlet-fidelity}
\end{equation}
Here $\ket{\psi_{\mathrm{ref}}}$ is the reference state 
\begin{align}
    \ket{\psi_{\mathrm{ref}}(T)} = U_{\mathrm{Trot}} \ket{\psi_0}
\end{align}
The denominator in Eq. \ref{eq:conditional-singlet-fidelity} is unity, and conditional fidelity (up to numerical precision) captures synthesis fidelity.   The accompanying
$T$-count comparison then answers a different question from the leakage plot:
conditional fidelity tests whether the component that remains in the physical
sector has the correct dynamics, while leakage measures how much probability
left that sector in the first place.  Together, these diagnostics show that the
advantage of symmetry-preserving compilation need not appear as a larger raw
state fidelity.  Rather, it is the ability to reproduce the physical in-sector
dynamics with fewer $T$ gates while preventing synthesis error from populating
forbidden total-spin sectors.

\begin{figure*}[t]
    \centering
    \includegraphics[width=0.48\textwidth]{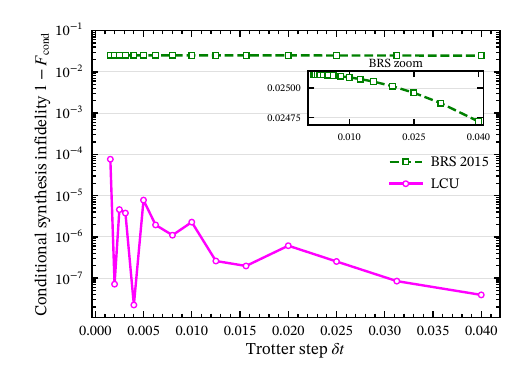}
    \hfill
    \includegraphics[width=0.48\textwidth]{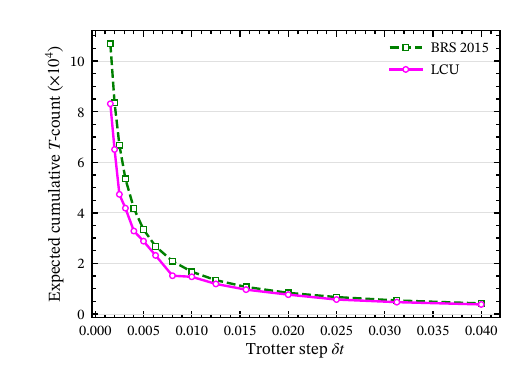}
    \caption{Accuracy and cost within the physical $S=0$ sector with
    synthesis tolerances $\epsilon_{\mathrm{LCU}}=2\times10^{-5}$ and
    $\epsilon_{\mathrm{BRS}}=2\times10^{-3}$; these precision values were chosen to match T counts.  \textbf{(a)} Conditional
    synthesis infidelity after projection onto the singlet sector and
    renormalization.  \textbf{(b)} Expected total $T$-count for the same
    BRS 2015 and LCU circuits.  At these method-specific tolerances, the LCU
    method has significantly lower conditional synthesis infidelity over the full step-size range.  The initial state is
    $\ket{S}_{01}\otimes\ket{S}_{23}$ with $N=4$ and $t_{tot}=1$.
}
    \label{fig:conditional-singlet-fidelity-cost}
\end{figure*}

\section{Extensions beyond the Heisenberg model}
\label{sec:extensions}

\subsection{Relation to commutants}

In this work, we consider the isotropic 1D Heisenberg Hamiltonian
on $\mathcal{H} = (\mathbb{C}^2)^{\otimes N}$, the Hilbert space
of $N$ spin-$1/2$ particles. Each nearest-neighbor interaction
can be expressed in terms of the swap operator $P_{i,i+1}$
and the identity. These swaps generate all permutations of the
spins, whose complex linear combinations form the permutation
algebra
\begin{equation}
\mathcal{A}_{S}
= \operatorname{span}_{\mathbb{C}}
\{P_{\pi} \mid \pi \in S_N\}.
\end{equation}
Here, $P_{\pi}$ permutes the tensor factors according to $\pi$,
and $\mathcal{A}_{S}$ is the image of the group algebra
$\mathbb{C}[S_N]$ acting on $\mathcal{H}$.

Global spin rotations act on the same Hilbert space as
$U^{\otimes N}$, with $U \in SU(2)$. Let $\mathcal{A}_{U}$
denote the complex unital algebra generated by these rotations,
or equivalently by the three total spin components. Permutations
commute with global rotations because applying the same rotation
to every spin is independent of their ordering.
Schur--Weyl duality strengthens this observation by establishing
that these two algebras are each other's
commutants~\cite{Bacon2006Schur}:
\begin{equation}
\operatorname{Comm}(\mathcal{A}_{S}) = \mathcal{A}_{U},
\qquad
\operatorname{Comm}(\mathcal{A}_{U}) = \mathcal{A}_{S}.
\end{equation}
The commutant of an algebra consists of all operators on
$\mathcal{H}$ that commute with every element of that algebra.
Consequently, any Hamiltonian invariant under global $SU(2)$
rotations belongs to $\mathcal{A}_{S}$ and can be written as
\begin{equation}
H = \sum_{\pi \in S_N} c_{\pi} P_{\pi}.
\end{equation}
This representation applies to arbitrary globally
$SU(2)$-invariant spin-$1/2$ interactions, including longer-range
and multi-spin terms. It does not require $H$ itself to be
invariant under permutations.

Expressing a Hamiltonian in terms of permutations does not,
however, immediately give a circuit for its time evolution.
Although every permutation can be decomposed into a product
of two-site swaps, including nearest-neighbor swaps on a
connected chain, a general Hamiltonian need not be a linear
combination of individual swaps. Moreover, a product
decomposition of a permutation does not directly provide a
decomposition of the exponential of a Hamiltonian containing
that permutation. In general, fractional-swap gates
$e^{-i\theta P_{ab}}$ acting on the system alone cannot generate
every globally $SU(2)$-invariant unitary. These restrictions
can be overcome using two ancillary
qubits~\cite{Marvian_2024}.

The two-site spin-$1/2$ case is particularly simple. A pair
of spins decomposes into a spin-0 singlet and a spin-1 triplet,
each occurring once. Thus, a Hamiltonian term $h_{ab}$
supported only on this pair and invariant under their joint
rotations has the form
\begin{equation}
h_{ab}
= a_0 \Pi_0 + a_1 \Pi_1
= \frac{a_0+a_1}{2} I
+ \frac{a_1-a_0}{2} P_{ab},
\end{equation}
where $\Pi_0$ and $\Pi_1$ project onto the singlet and triplet
sectors, respectively. Its time evolution is therefore a
fractional swap up to an overall phase. This is the local
structure used to compile the Heisenberg interactions.

If a term also acts on other spins but commutes with all three
components of the pair's total spin,
$\mathbf{J}_{ab}=\mathbf{S}_a+\mathbf{S}_b$, the more general
form is
\begin{align}
[h,J_{ab}^{\alpha}]=0
\quad \text{for } \alpha=x,y,z
\quad \\ \Longrightarrow 
\quad
h = \Pi_0 \otimes B_0 + \Pi_1 \otimes B_1,
\end{align}
where $B_0$ and $B_1$ act on the remaining spins in its support.
Such a term need not reduce to a fractional swap on the pair.
This condition is also stronger than global $SU(2)$ invariance:
a globally invariant multi-spin term need not commute with
rotations of a selected pair alone.

For higher physical spins, even a two-site interaction has
more total-spin sectors. Two spin-1 particles, for example,
have sectors $j=0,1,2$, and the AKLT bond interaction is
proportional to the projector onto $j=2$. A physical swap
cannot distinguish the $j=0$ and $j=2$ sectors, since both
are symmetric under exchange. The simple reduction to the
identity and a swap therefore no longer applies. Encoding
each spin-1 particle into two virtual spin-$1/2$ particles
also requires a compilation that respects the encoded
spin-1 subspace.

Nevertheless, some interactions beyond the Heisenberg model
admit a compilation using fractional swaps. One example is
the scalar spin chirality,
\begin{equation}
\chi = \mathbf{S}_1 \cdot
(\mathbf{S}_2 \times \mathbf{S}_3),
\end{equation}
a globally $SU(2)$-invariant three-spin interaction. For
spin-$1/2$ particles, with $\hbar=1$, it satisfies
\begin{equation}
\chi = \frac{i}{4}
\left(P_{12}P_{23}-P_{23}P_{12}\right).
\end{equation}
Its time evolution can be compiled using fractional swaps,
with every gate preserving global $SU(2)$ symmetry exactly, while incurring finite Trotter order error. 
See Appendix~\ref{sec:chirality} for the construction.

\subsection{Application to lattice gauge theory}
Digital lattice-gauge simulation protocols can preserve gauge invariance by constructing each stroboscopic interaction from gauge-invariant operations, including ancilla-mediated implementations of multibody terms~\cite{Zohar2017DigitalLGT}. To examine how fault-tolerant synthesis errors can compromise this protection, we consider the 1D $Z_2$ gauge-Higgs model~\cite{Gaugehiggs}, which describes a discretized version of matter coupled lattice gauge theory:
\begin{align}
    H= -J\sum_{j} \tau_j^z \sigma_{j,j+1}^z\tau_{j+1}^z -h_g\sum_{j} \sigma_{j,j+1}^x \\-h_m\sum_{j} \tau_j^x.
\end{align}
Here $\tau_j^z$ lives on sites (the matter field), while $\sigma_{j,j+1}^z$ lives on links (the gauge field). A single $H_j$ term is
\begin{align*}
    H_j &= H_{j,A}+H_{j,B}+H_{j,C}\\
        &= -J \tau_j^z \sigma_{j,j+1}^z\tau_{j+1}^z -h_g \sigma_{j,j+1}^x -h_m \tau_j^x.
\end{align*}
The Gauss-law constraint is
\begin{align}
    G_j = \sigma_{j-1,j}^x \tau_{j}^x \sigma_{j,j+1}^x.
\end{align}
Each term commutes with $G_j$: the second and third terms are already Pauli-$X$ terms, while the first touches the Gauss-law constraint on two sites and therefore also commutes. Since each term individually commutes with the Gauss-law constraint, we can separate them while incurring a gauge-preserving Trotter error. A single Trotter step can thus be written as
\begin{align}
   e^{-iH_j\delta t} = e^{-iH_{j,A}\delta t} e^{-iH_{j,B}\delta t} e^{-iH_{j,C}\delta t} +O(\delta t^2).
\end{align}
Each term can be generated by conjugating a single rotation gate
$R_z(\theta)=e^{-i\theta Z/2}$ with Clifford gates. Up to a global
phase, a generic unitary approximant sufficiently close to this target
can be written as
\begin{align}
    \widetilde R_z(\theta)
    =\exp\!\left[-i\left(
        \left(\frac{\theta}{2}+\delta_z\right)Z
        +\delta_x X+\delta_y Y
    \right)\right],
    \label{eq:generic-off-axis-rz}
\end{align}
where $\delta_z$ is an axial angle error and $\delta_x,\delta_y$ are
transverse errors. Standard symmetry-agnostic synthesis constrains the
distance to the target unitary but does not, in general, force the
transverse coefficients to vanish.

For example, the matter-field term $H_{j,C}=-h_m\tau_j^x$ is obtained
by conjugating a $Z$ rotation with a Hadamard. Using
$HZH=X$, $HXH=Z$, and $HYH=-Y$, Eq.~\eqref{eq:generic-off-axis-rz}
gives
\begin{align}
    H\widetilde R_z(\theta)H
    &=\exp\!\left[-i\widetilde h_j\right], \nonumber\\
    \widetilde h_j
    &=\left(\frac{\theta}{2}+\delta_z\right)\tau_j^x
      +\delta_x\tau_j^z-\delta_y\tau_j^y.
    \label{eq:trotter-with-synthesis}
\end{align}
The intended generator $\tau_j^x$, including the axial over- or
under-rotation $\delta_z\tau_j^x$, commutes with the Gauss-law operator
$G_j$. By contrast,
$[\tau_j^z,G_j]\neq0$ and $[\tau_j^y,G_j]\neq0$, so nonzero
$\delta_x$ or $\delta_y$ generically breaks gauge invariance. The same
reasoning applies to the Clifford conjugations used for the other Pauli
terms: an axial synthesis error changes only the coefficient of the
intended gauge-invariant generator, whereas transverse errors introduce
different Pauli strings that need not commute with the Gauss-law
operators. The BRS 2015 construction is axial, so its successful branch
implements $R_z(\theta')$ for a nearby angle $\theta'$; after the same
Clifford conjugation, it therefore changes the coupling strength without
breaking gauge invariance.

\subsection{Fermionic models: Fermi--Hubbard and electronic structure}

The construction above extends to fermionic models, whose continuous symmetries include U(1) particle-number conservation, meaning that the dynamics conserves a quantity generated by a sum of local $Z$ operators after mapping to qubits. A central example is the Fermi--Hubbard model,
\begin{equation}
H = -t\sum_{\langle i,j\rangle,\sigma}\left(c^\dagger_{i\sigma}c_{j\sigma}+\mathrm{h.c.}\right) + U\sum_i n_{i\uparrow}n_{i\downarrow},
\end{equation}
where $c_{i\sigma}$ annihilates a fermion with spin $\sigma$ on site $i$, $n_{i\sigma}=c^\dagger_{i\sigma}c_{i\sigma}$, $t$ is the hopping amplitude and $U$ the on-site interaction. The model conserves the number of fermions of each spin species, $N_\sigma=\sum_i n_{i\sigma}$, and is in addition invariant under global spin rotations generated by $\mathbf{S}=\frac{1}{2}\sum_i c^\dagger_{i}\boldsymbol{\sigma}c_{i}$, where $c_i=(c_{i\uparrow},c_{i\downarrow})^T$ and $\boldsymbol{\sigma}$ is the vector of Pauli matrices. To represent it on qubits we use the Jordan--Wigner transformation, which assigns one qubit to each fermionic mode with $n=(1-Z)/2$ and attaches strings of $Z$ operators to reproduce fermionic anticommutation~\cite{Jordan1928}. Terms acting on modes that are adjacent in the Jordan--Wigner ordering then become two-qubit gates: hopping gates $e^{i\theta(X_aX_b+Y_aY_b)/2}$ with $\theta=t\,\delta t$, and interaction gates $e^{-iU\delta t\, n_an_b}=\mathrm{\cphase}(-U\delta t)$.

The interaction gate is a $\cphase$ gate and can be compiled with the BRS 2015 or LCU constructions of Sec.~V. The hopping gate is diagonalized by the basis change $B$ of Eq.~(21): $(XX+YY)/2$ has eigenvalues $0,0,+1,-1$ on $\ket{00},\ket{11},\ket{T_0},\ket{S}$, so that $e^{i\theta(X_aX_b+Y_aY_b)/2}=BDB^\dagger$ with $D=\mathrm{diag}(1,1,e^{i\theta},e^{-i\theta})$. We write $D=e^{-i\alpha Z_b}e^{i\beta Z_aZ_b}$ with $\alpha=\beta=-\theta/2$, and compile both factors with an axial method. If the implemented angles $\alpha'$ and $\beta'$ carry independent synthesis errors, the compiled gate is
\begin{equation}
BD'B^\dagger = e^{i\theta'(X_aX_b+Y_aY_b)/2}\, e^{i(\beta'-\alpha')(Z_a+Z_b)/2}
\end{equation}
with $\theta'=-(\alpha'+\beta')$. The synthesis error therefore rescales the hopping amplitude and adds a bond-dependent shift of the chemical potential, $Z_a+Z_b=2-2(n_a+n_b)$. Both terms conserve $N_\uparrow$ and $N_\downarrow$ exactly. If, moreover, the hopping gates of the two spin species on a given bond use the same Clifford+T approximant, the error terms are proportional to the spin-summed hopping and density, which are spin singlets. The compiled Trotter step then commutes with $\mathbf{S}$ and conserves total spin $\mathbf{S}^2$ exactly, in the same way as in the Heisenberg model. This is possible because the angle implemented by the BRS 2015 and LCU constructions is fixed by the chosen triple $(z,y,l)$ and is independent of the measurement history. The default symmetry-agnostic synthesis, by contrast, produces off-axis $X$ and $Y$ components, which change the particle number and populate sectors with the wrong filling.

The same structure applies to the electronic structure Hamiltonian of quantum chemistry, $H=\sum_{pq}h_{pq}c^\dagger_pc_q+\frac{1}{2}\sum_{pqrs}h_{pqrs}c^\dagger_pc^\dagger_qc_rc_s$, where $c_p$ annihilates an electron in spin orbital $p$. Its two-body terms act on four modes, but in a double-factorized representation the Hamiltonian is written as a sum of orbital rotations $\mathcal{U}_\ell$, which are unitary changes of the single-particle basis, surrounding density-density terms $\sum_{pq}\lambda^{(\ell)}_{pq}n_pn_q$~\cite{Motta2021}. Each orbital rotation decomposes into Givens rotations $e^{\theta(c^\dagger_pc_q-c^\dagger_qc_p)}$, which act on adjacent modes as $e^{i\theta(X_aY_b-Y_aX_b)/2}$. This gate is diagonalized by $B$ followed by a phase gate $S$ on one qubit, which is again an exact Clifford+T basis change, and its diagonal part is compiled axially as above. The density-density terms are products of CPHASE gates and $R_z$ rotations. The compiled Trotter step therefore conserves particle number and $S^z$ exactly. In the spin-restricted case, where the orbitals are the same for both spin species, each orbital rotation is generated by the spin-summed operators $E_{pq}=\sum_\sigma c^\dagger_{p\sigma}c_{q\sigma}$, $\mathcal{U}_\ell=\exp\big[\sum_{p<q}\kappa^{(\ell)}_{pq}(E_{pq}-E_{qp})\big]$ with real $\kappa^{(\ell)}_{pq}$, and each density-density term involves the spin-summed densities $n_p=n_{p\uparrow}+n_{p\downarrow}$. Both $E_{pq}$ and $n_p$ are spin singlets, so the exact Trotter step commutes with $\mathbf{S}$. After decomposition, every Givens rotation on the spatial orbitals $(p,q)$ appears as a pair $G^\uparrow_{pq}(\theta)G^\downarrow_{pq}(\theta)$ acting on the two spin species, and every coefficient $\lambda^{(\ell)}_{pq}$ appears in the four CPHASE gates on the mode pairs $(p\sigma,q\sigma')$, $\sigma,\sigma'\in\{\uparrow,\downarrow\}$, together with $R_z$ rotations on $p\uparrow$ and $p\downarrow$. We compile all gates within each such group with the same Clifford+T approximant. The implemented Givens pair is then $\exp[\theta'(E_{pq}-E_{qp})]$ multiplied by a chemical-potential shift proportional to $n_p+n_q$, the implemented density-density term is $\lambda'^{(\ell)}_{pq}n_pn_q$, and the $R_z$ errors shift the coefficient of $n_p$. All of these are spin singlets, so the compiled Trotter step conserves $\mathbf{S}^2$ exactly. If instead the gates within a group were compiled independently, the unequal angles would generate terms such as $n_{p\uparrow}n_{q\uparrow}+n_{p\downarrow}n_{q\downarrow}$, which contain $S^z_pS^z_q$ and break the SU(2) symmetry.

\section{Discussion and conclusion}
In this work, we have shown that Trotterized time evolution compiled into a discrete fault-tolerant gate set does not necessarily preserve continuous symmetries such as the $SU(2)$ symmetry of the Heisenberg model. However, we can use an ancilla measurement to implement a repeat-until-success operation that does preserve $SU(2)$ symmetry. To this end, we use BRS 2015 synthesis method to compile a controlled-phase gate with over- and under-rotations about a fixed axis, thereby preserving $SU(2)$ symmetry.

We also introduce a measurement-assisted LCU construction that compiles the
controlled-phase gate directly. We compare this construction with the
symmetry-preserving BRS 2015 implementation, which realizes the same gate
using three axial single-qubit rotations. Over the parameter range studied,
the direct LCU construction has a lower expected \(T\)-count than the BRS
2015 construction. Both methods preserve \(SU(2)\) symmetry exactly because
their synthesis errors modify only the relative phase between the singlet
and triplet sectors, without coupling those sectors.

We further show that coherent angle errors can be reduced by alternating
approximants that over- and under-rotate relative to the target angle. This
alternating synthesis schedule cancels the leading signed-angle error
pairwise and can improve the final-state fidelity. 
Finally, we apply these symmetry-preserving compilation methods to the
extraction of the Kardar--Parisi--Zhang transport exponent in the Heisenberg
model and find that they can reach a given transport accuracy with a lower
expected \(T\)-count over the parameter range considered.

This study opens the door to further exploration of symmetry-preserving methods in other models with continuous symmetries and of the effects of symmetry-breaking errors on the dynamics of such systems. We currently consider spin-$1/2$ systems, but this problem would be especially interesting for higher-spin systems, where the Trotter decomposition is less straightforward and spin operators are not unitary. Symmetry-preserving compilation may also be useful in lattice gauge theories, where an appropriate compilation can preserve gauge symmetries without penalty terms. We hope that this work motivates further research on symmetry-preserving quantum compilation.
\section*{Acknowledgments}
We thank Zohreh Davoudi, Nazli Koyluoglu, and Yuxin Wang for helpful discussions and questions. We also acknowledge the use of ChatGPT 5.6 in calculations, writing some sections of code, and in grammatical checks of the manuscript. We have verified that these calculations and code is correct. M.M. and D.G. are supported by the RQS QLCI grant OMA-2120757. Z.Y.W.~was supported in part by the NSF STAQ program (awards No.~1818914 and No.~2325080).
    \bibliography{ref}

@article{Feynman1982,
  author = {Feynman, Richard P.},
  title = {Simulating Physics with Computers},
  journal = {International Journal of Theoretical Physics},
  volume = {21}, pages = {467--488}, year = {1982},
  doi = {10.1007/BF02650179}
}

@article{Marvian_2024,
   title={Rotationally Invariant Circuits: Universality with the Exchange Interaction and Two Ancilla Qubits},
   volume={132},
   ISSN={1079-7114},
   url={http://dx.doi.org/10.1103/PhysRevLett.132.130201},
   DOI={10.1103/physrevlett.132.130201},
   number={13},
   journal={Physical Review Letters},
   publisher={American Physical Society (APS)},
   author={Marvian, Iman and Liu, Hanqing and Hulse, Austin},
   year={2024},
   month=Mar }

@article{Lloyd1996,
  author = {Lloyd, Seth}, title = {Universal Quantum Simulators},
  journal = {Science}, volume = {273}, number = {5278},
  pages = {1073--1078}, year = {1996},
  doi = {10.1126/science.273.5278.1073}
}

@article{Childs2021Trotter,
  author = {Childs, Andrew M. and Su, Yuan and Tran, Minh C. and Wiebe, Nathan and Zhu, Shuchen},
  title = {Theory of {Trotter} Error with Commutator Scaling},
  journal = {Physical Review X}, volume = {11}, pages = {011020}, year = {2021},
  doi = {10.1103/PhysRevX.11.011020}
}

@article{Bethe1931,
  author = {Bethe, Hans},
  title = {Zur Theorie der Metalle. {I}. Eigenwerte und Eigenfunktionen der linearen Atomkette},
  journal = {Zeitschrift f{\"u}r Physik}, volume = {71}, pages = {205--226}, year = {1931},
  doi = {10.1007/BF01341708}
}

@article{DawsonNielsen2006,
  author = {Dawson, Christopher M. and Nielsen, Michael A.},
  title = {The {Solovay--Kitaev} Algorithm},
  journal = {Quantum Information and Computation}, volume = {6}, number = {1},
  pages = {81--95}, year = {2006}, doi = {10.26421/QIC6.1-6}
}

@article{Jordan1928,
  author  = {Jordan, P. and Wigner, E.},
  title   = {{\"U}ber das Paulische {\"A}quivalenzverbot},
  journal = {Zeitschrift f{\"u}r Physik},
  volume  = {47},
  pages   = {631--651},
  year    = {1928},
  doi     = {10.1007/BF01331938}
}

@article{Motta2021,
  author  = {Motta, Mario and Ye, Erika and McClean, Jarrod R. and Li, Zhendong and Minnich, Austin J. and Babbush, Ryan and Chan, Garnet Kin-Lic},
  title   = {Low rank representations for quantum simulation of electronic structure},
  journal = {npj Quantum Information},
  volume  = {7},
  pages   = {83},
  year    = {2021},
  doi     = {10.1038/s41534-021-00416-z}
}

@article{RossSelinger2016,
  author = {Ross, Neil J. and Selinger, Peter},
  title = {Optimal Ancilla-Free {Clifford+$T$} Approximation of $z$-Rotations},
  journal = {Quantum Information and Computation}, volume = {16}, number = {11--12},
  pages = {901--953}, year = {2016}, doi = {10.26421/QIC16.11-12-1}
}

@article{Kliuchnikov2013Exact,
  author = {Kliuchnikov, Vadym and Maslov, Dmitri and Mosca, Michele},
  title = {Fast and Efficient Exact Synthesis of Single-Qubit Unitaries Generated by {Clifford} and {$T$} Gates},
  journal = {Quantum Information and Computation}, volume = {13}, number = {7--8},
  pages = {607--630}, year = {2013}, eprint = {1206.5236},
  archivePrefix = {arXiv}, primaryClass = {quant-ph}
}

@article{GilesSelinger2013,
  author = {Giles, Brett and Selinger, Peter},
  title = {Exact Synthesis of Multiqubit {Clifford+$T$} Circuits},
  journal = {Physical Review A}, volume = {87}, pages = {032332}, year = {2013},
  doi = {10.1103/PhysRevA.87.032332}
}

@article{Bocharov2015RUS,
  author = {Bocharov, Alex and Roetteler, Martin and Svore, Krysta M.},
  title = {Efficient Synthesis of Universal Repeat-Until-Success Quantum Circuits},
  journal = {Physical Review Letters}, volume = {114}, pages = {080502}, year = {2015},
  doi = {10.1103/PhysRevLett.114.080502}
}

@article{Manousakis1991,
  author  = {Manousakis, Efstratios},
  title   = {The spin-1/2 {H}eisenberg antiferromagnet on a square
             lattice and its application to the cuprous oxides},
  journal = {Reviews of Modern Physics},
  volume  = {63},
  number  = {1},
  pages   = {1--62},
  year    = {1991},
  doi     = {10.1103/RevModPhys.63.1}
}

@article{Amico2008,
  author  = {Amico, Luigi and Fazio, Rosario and Osterloh, Andreas
             and Vedral, Vlatko},
  title   = {Entanglement in many-body systems},
  journal = {Reviews of Modern Physics},
  volume  = {80},
  number  = {2},
  pages   = {517--576},
  year    = {2008},
  doi     = {10.1103/RevModPhys.80.517}
}

@article{Piroli2016,
  author  = {Piroli, Lorenzo and Vernier, Eric and Calabrese, Pasquale},
  title   = {Exact steady states for quantum quenches in integrable
             {H}eisenberg spin chains},
  journal = {Physical Review B},
  volume  = {94},
  number  = {5},
  pages   = {054313},
  year    = {2016},
  doi     = {10.1103/PhysRevB.94.054313}
}

@article{White1992,
  author  = {White, Steven R.},
  title   = {Density Matrix Formulation for Quantum
             Renormalization Groups},
  journal = {Physical Review Letters},
  volume  = {69},
  number  = {19},
  pages   = {2863--2866},
  year    = {1992},
  doi     = {10.1103/PhysRevLett.69.2863}
}

@article{Childs2018,
  author  = {Childs, Andrew M. and Maslov, Dmitri and Nam, Yunseong
             and Ross, Neil J. and Su, Yuan},
  title   = {Toward the First Quantum Simulation with Quantum Speedup},
  journal = {Proceedings of the National Academy of Sciences},
  volume  = {115},
  number  = {38},
  pages   = {9456--9461},
  year    = {2018},
  doi     = {10.1073/pnas.1801723115}
}

@article{PaetznickSvore2014,
  author = {Paetznick, Adam and Svore, Krysta M.},
  title = {Repeat-Until-Success: Non-Deterministic Decomposition of Single-Qubit Unitaries},
  journal = {Quantum Information and Computation}, volume = {14}, number = {15--16},
  pages = {1277--1301}, year = {2014}, eprint = {1311.1074},
  archivePrefix = {arXiv}, primaryClass = {quant-ph}
}

@book{GeometricIntegrationBook,
  title={Geometric Numerical Integration: Structure-Preserving Algorithms for Ordinary Differential Equations},
  author={Hairer, Ernst and Lubich, Christian and Wanner, Gerhard},
  year={2006},
  edition={2nd},
  publisher={Springer-Verlag},
  address={Berlin, Heidelberg},
  isbn={978-3-540-30663-4},
  doi={10.1007/3-540-30666-8}
}

@article{WisdomSymplectic,
       author = {{Wisdom}, Jack and {Holman}, Matthew},
        title = "{Symplectic maps for the N-body problem.}",
      journal = {The Astronomical Journal},
         year = 1991,
        month = oct,
       volume = {102},
        pages = {1528-1538},
          doi = {10.1086/115978},
       adsurl = {https://ui.adsabs.harvard.edu/abs/1991AJ....102.1528W}
}

@article{SinghTensorNetwork,
  title = {Tensor network decompositions in the presence of a global symmetry},
  author = {Singh, Sukhwinder and Pfeifer, Robert N. C. and Vidal, Guifr\'e},
  journal = {Phys. Rev. A},
  volume = {82},
  issue = {5},
  pages = {050301(R)},
  numpages = {4},
  year = {2010},
  month = {Nov},
  publisher = {American Physical Society},
  doi = {10.1103/PhysRevA.82.050301},
  url = {https://link.aps.org/doi/10.1103/PhysRevA.82.050301}
}

@article{SinghTensorNetwork2,
  title = {Tensor network states and algorithms in the presence of a global U(1) symmetry},
  author = {Singh, Sukhwinder and Pfeifer, Robert N. C. and Vidal, Guifre},
  journal = {Phys. Rev. B},
  volume = {83},
  issue = {11},
  pages = {115125},
  numpages = {22},
  year = {2011},
  month = {Mar},
  publisher = {American Physical Society},
  doi = {10.1103/PhysRevB.83.115125},
  url = {https://link.aps.org/doi/10.1103/PhysRevB.83.115125}
}

@article{Berry2015LCU,
  author = {Berry, Dominic W. and Childs, Andrew M. and Cleve, Richard and Kothari, Robin and Somma, Rolando D.},
  title = {Simulating {Hamiltonian} Dynamics with a Truncated {Taylor} Series},
  journal = {Physical Review Letters}, volume = {114}, pages = {090502}, year = {2015},
  doi = {10.1103/PhysRevLett.114.090502}
}

@misc{Bacon2006Schur,
  author = {Bacon, Dave and Chuang, Isaac L. and Harrow, Aram W.},
  title = {The Quantum {Schur} Transform: {I}. Efficient Qudit Circuits},
  year = {2006}, eprint = {quant-ph/0601001}, archivePrefix = {arXiv},
  primaryClass = {quant-ph}
}

@article{DiVincenzo2000Exchange,
  author = {DiVincenzo, David P. and Bacon, Dave and Kempe, Julia and Burkard, Guido and Whaley, K. Birgitta},
  title = {Universal Quantum Computation with the Exchange Interaction},
  journal = {Nature}, volume = {408}, pages = {339--342}, year = {2000},
  doi = {10.1038/35042541}
}

@article{LaBordeWilde2022,
  author = {LaBorde, Margarite L. and Wilde, Mark M.},
  title = {Quantum Algorithms for Testing {Hamiltonian} Symmetry},
  journal = {Physical Review Letters}, volume = {129}, pages = {160503}, year = {2022},
  doi = {10.1103/PhysRevLett.129.160503}
}

@article{DeNardis2020Solitons,
  author = {De Nardis, Jacopo and Gopalakrishnan, Sarang and Ilievski, Enej and Vasseur, Romain},
  title = {Superdiffusion from Emergent Classical Solitons in Quantum Spin Chains},
  journal = {Physical Review Letters}, volume = {125}, pages = {070601}, year = {2020},
  doi = {10.1103/PhysRevLett.125.070601}
}

@article{DeNardis2021Stability,
  author = {De Nardis, Jacopo and Gopalakrishnan, Sarang and Vasseur, Romain and Ware, Brayden},
  title = {Stability of Superdiffusion in Nearly Integrable Spin Chains},
  journal = {Physical Review Letters}, volume = {127}, pages = {057201}, year = {2021},
  doi = {10.1103/PhysRevLett.127.057201}
}

@misc{Amy2024Cyclotomic,
  author = {Amy, Matthew and Glaudell, Andrew N. and Kelso, Shaun and Maxwell, William and Mendelson, Samuel S. and Ross, Neil J.},
  title = {Exact Synthesis of Multiqubit {Clifford}-Cyclotomic Circuits},
  year = {2024},
  eprint = {2311.07741},
  archivePrefix = {arXiv},
  primaryClass = {quant-ph}
}

@article{Barenco1995Elementary,
  author = {Barenco, Adriano and Bennett, Charles H. and Cleve, Richard and DiVincenzo, David P. and Margolus, Norman and Shor, Peter and Sleator, Tycho and Smolin, John A. and Weinfurter, Harald},
  title = {Elementary Gates for Quantum Computation},
  journal = {Physical Review A},
  volume = {52},
  pages = {3457--3467},
  year = {1995},
  doi = {10.1103/PhysRevA.52.3457}
}

@misc{Kliuchnikov2013UnitarySynthesis,
  author = {Kliuchnikov, Vadym},
  title = {Synthesis of Unitaries with {Clifford+$T$} Circuits},
  year = {2013},
  eprint = {1306.3200},
  archivePrefix = {arXiv},
  primaryClass = {quant-ph}
}

@article{Zohar2017DigitalLGT,
  author = {Zohar, Erez and Farace, Alessandro and Reznik, Benni and Cirac, J. Ignacio},
  title = {Digital Lattice Gauge Theories},
  journal = {Physical Review A},
  volume = {95},
  pages = {023604},
  year = {2017},
  doi = {10.1103/PhysRevA.95.023604}
}

@article{Freedman2021SymmetryProtected,
  author = {Freedman, Michael H. and Hastings, Matthew B. and Shokrian Zini, Modjtaba},
  title = {Symmetry Protected Quantum Computation},
  journal = {Quantum},
  volume = {5},
  pages = {554},
  year = {2021},
  doi = {10.22331/q-2021-09-28-554}
}

@article{Selinger2013TDepth,
  author = {Selinger, Peter},
  title = {Quantum Circuits of {$T$}-Depth One},
  journal = {Physical Review A},
  volume = {87},
  pages = {042302},
  year = {2013},
  doi = {10.1103/PhysRevA.87.042302}
}

@misc{NegishiYang2026LocalSymmetry,
  author = {Negishi, Naoki and Yang, Bo},
  title = {Beyond Commutativity: Redesigning {Trotter} Decomposition via Local Symmetry},
  year = {2026},
  eprint = {2605.16016},
  archivePrefix = {arXiv},
  primaryClass = {quant-ph}
}

@article{Prosen_theory,
  title = {Kardar-Parisi-Zhang Physics in the Quantum Heisenberg Magnet},
  author = {Ljubotina, Marko and \ifmmode \check{Z}\else \v{Z}\fi{}nidari\ifmmode \check{c}\else \v{c}\fi{}, Marko and Prosen, Toma\ifmmode \check{z}\else \v{z}\fi{}},
  journal = {Phys. Rev. Lett.},
  volume = {122},
  issue = {21},
  pages = {210602},
  numpages = {6},
  year = {2019},
  month = {May},
  publisher = {American Physical Society},
  doi = {10.1103/PhysRevLett.122.210602},
  url = {https://link.aps.org/doi/10.1103/PhysRevLett.122.210602}
}

@article{
GoogleKPZ,
author = {E. Rosenberg  and T. I. Andersen  and R. Samajdar  and A. Petukhov  and V. Khemani  and S. Gopalakrishnan  and T. Prosen  and P. Roushan },
title = {Dynamics of magnetization at infinite temperature in a Heisenberg spin chain},
journal = {Science},
volume = {384},
number = {6691},
pages = {48-53},
year = {2024},
doi = {10.1126/science.adi7877},
URL = {https://www.science.org/doi/abs/10.1126/science.adi7877},
eprint = {https://www.science.org/doi/pdf/10.1126/science.adi7877}}

@misc{matsumoto2008representationquantumcircuitsclifford,
      title={Representation of Quantum Circuits with Clifford and $\pi/8$ Gates}, 
      author={Ken Matsumoto and Kazuyuki Amano},
      year={2008},
      eprint={0806.3834},
      archivePrefix={arXiv},
      primaryClass={quant-ph},
      url={https://arxiv.org/abs/0806.3834}, 
}

@article{Banerjee_2026,
   title={Digital Quantum Simulation of Spin Transport},
   volume={136},
   ISSN={1079-7114},
   url={http://dx.doi.org/10.1103/mx9k-kdlj},
   DOI={10.1103/mx9k-kdlj},
   number={5},
   journal={Physical Review Letters},
   publisher={American Physical Society (APS)},
   author={Lee, Yi-Ting and Pokharel, Bibek and Cohn, Jeffrey and Schleife, André and Banerjee, Arnab},
   year={2026},
   month=Feb }

@article{Gaugehiggs,
  title = {Digital quantum simulation and pseudoquantum simulation of the ${\mathbb{Z}}_{2}$ gauge--Higgs model},
  author = {Ding, Yiming and Cui, Xiaopeng and Shi, Yu},
  journal = {Phys. Rev. D},
  volume = {105},
  issue = {5},
  pages = {054508},
  numpages = {13},
  year = {2022},
  month = {Mar},
  publisher = {American Physical Society},
  doi = {10.1103/PhysRevD.105.054508},
  url = {https://link.aps.org/doi/10.1103/PhysRevD.105.054508}
}

@article{Bloch_group_expt,
   title={Kardar-Parisi-Zhang universality in discrete two-dimensional driven-dissipative exciton polariton condensates},
   volume={4},
   ISSN={2643-1564},
   url={http://dx.doi.org/10.1103/PhysRevResearch.4.043207},
   DOI={10.1103/physrevresearch.4.043207},
   number={4},
   journal={Physical Review Research},
   publisher={American Physical Society (APS)},
   author={Deligiannis, Konstantinos and Fontaine, Quentin and Squizzato, Davide and Richard, Maxime and Ravets, Sylvain and Bloch, Jacqueline and Minguzzi, Anna and Canet, Léonie},
   year={2022},
   month=Dec }

@misc{gosset_tcount,
  title        = {An Algorithm for the T-count},
  author       = {Gosset, David and Kliuchnikov, Vadym and Mosca, Michele and Russo, Vincent},
  year         = {2013},
  eprint       = {1308.4134},
  archivePrefix= {arXiv},
  primaryClass = {quant-ph}
}

\appendix
\section{Compilation of roots of swaps in exact gates}\label{sec: swapexact}
We can show that certain roots can be compiled exactly in terms of Clifford+T.
\begin{itemize}
    \item $\swap$
    \begin{align}
        \swap = \cnott{1}{2}\cnott{2}{1}\cnott{1}{2}
    \end{align}
    \item $\sqrt{\swap}$
    \begin{align}
        \sqrt{\swap} = \cnott{1}{2} \h{2} \cphase{(\pi/2)} \h{2} \cnott{1}{2}
    \end{align}
    where $\cphase{(\pi/2)}$ can then be compiled in terms of $R_z(\pi/4)$ or T gates. 
    \item $\swap^{1/4}$
       \begin{align}
        {(\swap)}^{1/4} = \cnott{1}{2} \h{2} \cphase{(\pi/4)} \h{2} \cnott{1}{2}
    \end{align}
    $\cphase{(\pi/4)}$ is the controlled-$T$ (CT) gate, which can be compiled using an ancilla. A controlled-$T$ gate is given by
    \begin{align}
        CT\ket{a,b}\rightarrow e^{i\pi/4 a\wedge b}\ket{a,b}
    \end{align}
    \begin{itemize}
        \item Initialize ancilla to $\ket{0}$.
        \item Compute an AND gate:
        \begin{align}
            \text{AND}\ket{a,b,0}\rightarrow \ket{a,b,a\wedge b}
        \end{align}
        \item Apply a $T$ gate to the ancilla:
        \begin{align}
            T\ket{a,b,a\wedge b}=e^{i\pi/4 (a\wedge b)}\ket{a,b,a\wedge b}
        \end{align}
        \item Uncompute the ancilla using another AND operation implemented with a Toffoli gate:
        \begin{align}
            e^{i\pi/4 a\wedge b } \ket{a,b,a\wedge b}\rightarrow e^{i\pi/4 a\wedge b } \ket{a,b,0}
        \end{align}
    \end{itemize}
    An exact Toffoli has $T$-count seven in the unitary Clifford+$T$ accounting used in this work~\cite{gosset_tcount}; with four clean ancillas it can be implemented with $T$-depth one~\cite{Selinger2013TDepth}.

\end{itemize}
\section{Clifford+T is not sufficient for arbitrary SU(2) rotations}\label{sec:nogo}
We start with the simplifying assumption involving an element in the ring. Consider the unitary approximation of the form:
\begin{equation}
    \frac{a + i b\sqrt{2}}{(\sqrt{2})^k} = e^{i\phi}
\end{equation}
This implies that the norm is 1. Therefore:
\begin{equation}
    a^2 + 2b^2 = 2^k
\end{equation}
Here $a,b\in\mathbb{Z}$ and $k\in\mathbb{Z}_{\ge0}$.
If $b=0$, the norm equation gives $a^2=2^k$, and hence
$e^{i\phi}=\pm1$. If $a=0$, it gives $2b^2=2^k$,
and hence $e^{i\phi}=\pm i$. We therefore assume
$a,b\neq0$ in the following cases.

Let us define the parity of the integers $a$ and $b$ by factoring out powers of 2:
\begin{equation}
    a = a' 2^i, \quad b = b' 2^j
\end{equation}
where $a'$ and $b'$ are both odd integers. Substituting these into the norm equation:
\begin{equation}
    a'^2 2^{2i} + b'^2 2^{2j+1} = 2^k
\end{equation}

\textbf{Case 1: $2i < 2j+1$}
Factor out the smallest power of 2 ($2^{2i}$):
\begin{equation}
    a'^2 + b'^2 2^{2j+1-2i} = 2^{k-2i}
\end{equation}
Since $a'$ is odd, $a'^2$ is odd. Since $2j+1-2i \geq 1$, the term $b'^2 2^{2j+1-2i}$ is even.
\[ \text{LHS} = \text{Odd} + \text{Even} = \text{Odd} \]
However, the RHS ($2^{k-2i}$) is even (for $k-2i > 0$).
This is a contradiction. The left-hand side is an odd integer greater than one.
It therefore cannot equal a power of two, whose only
odd integer value is one. This is a contradiction.

\textbf{Case 2: $2i = 2j+1$}
This case is impossible because an even number cannot equal an odd number.

\textbf{Case 3: $2i > 2j+1$}
Factor out $2^{2j+1}$:
\begin{equation}
    a'^2 2^{2i - (2j+1)} + b'^2 = 2^{k-(2j+1)}
\end{equation}
Here, the first term is even and $b'^2$ is odd.
\[ \text{LHS} = \text{Even} + \text{Odd} = \text{Odd} \]
The RHS is a power of 2 (Even). The left-hand side is an odd integer greater than one.
It therefore cannot equal a power of two, whose only
odd integer value is one. This is a contradiction.

\textbf{General case:}
Now we move to the more complicated case involving full complex entries. The equation is generally of the form:
\begin{equation}
    \frac{a + b\sqrt{2}}{(\sqrt{2})^k} + i\, \frac{c + d\sqrt{2}}{(\sqrt{2})^k} = e^{i\phi}
\end{equation}
Multiplying by the complex conjugate to enforce the unitarity constraint (norm = 1):
\begin{equation}
    \frac{(a+b\sqrt{2})^2 + (c+d\sqrt{2})^2}{2^k} = 1
\end{equation}
Expanding the squares:
\begin{equation}
    (a^2 + 2b^2 + c^2 + 2d^2) + \sqrt{2}(2ab + 2cd) = 2^k
\end{equation}
For the RHS to be an integer ($2^k$), the irrational part must vanish:
\begin{equation}
    ab + cd = 0 \implies ab = -cd
\end{equation}
And the rational part must satisfy:
\begin{equation}
    a^2 + 2b^2 + c^2 + 2d^2 = 2^k
\end{equation}
\textbf{Parity analysis.}

Here $a,b,c,d\in\mathbb{Z}$ and
$k\in\mathbb{Z}_{\ge0}$.
We first suppose that all four coefficients are nonzero
and write
\begin{equation}
    a=2^i a', \qquad b=2^j b', \qquad
    c=2^\ell c', \qquad d=2^h d',
\end{equation}
where $a',b',c',d'$ are odd integers and
$i,j,\ell,h$ are nonnegative integers.
Since $ab=-cd$, the powers of two in these products
must agree:
\begin{equation}
    i+j=\ell+h.
\end{equation}

Suppose that $i\neq\ell$. The product relation then
implies $j\neq h$. Consequently, we can write
\begin{equation}
\begin{aligned}
    a^2+c^2 &= 2^{2\min(i,\ell)}A,\\
    2b^2+2d^2 &= 2^{1+2\min(j,h)}B,
\end{aligned}
\end{equation}
where $A$ and $B$ are positive odd integers.
The two exponents have opposite parity and therefore
cannot be equal. Factoring the smaller power of two
out of their sum leaves an odd integer greater than
one, contradicting
$a^2+c^2+2b^2+2d^2=2^k$.
Thus $i=\ell$, and the product relation also gives
$j=h$.

The norm equation now becomes
\begin{equation}
    2^{2i}(a'^2+c'^2)
    +2^{2j+1}(b'^2+d'^2)=2^k.
\end{equation}
Every odd square is congruent to $1$ modulo $4$,
so each sum in parentheses is congruent to $2$
modulo $4$. Hence
\begin{equation}
    a'^2+c'^2=2A,\qquad b'^2+d'^2=2B,
\end{equation}
for positive odd integers $A$ and $B$, giving
\begin{equation}
    2^{2i+1}A+2^{2j+2}B=2^k.
\end{equation}
Again, the two exponents have opposite parity.
Factoring out the smaller power of two leaves an
odd integer greater than one, which is impossible.
Therefore, all four coefficients cannot be nonzero.

\textbf{Cases with zero coefficients.}

If any coefficient vanishes, the relation $ab=-cd$
implies $ab=cd=0$. The possible remaining pairs give
one of the equations
\begin{equation}
\begin{aligned}
    a^2+c^2 &= 2^k,\\
    2(b^2+d^2) &= 2^k,\\
    a^2+2d^2 &= 2^k,\\
    c^2+2b^2 &= 2^k.
\end{aligned}
\end{equation}
The last two equations are of the form treated in
the initial example. They cannot have both
coefficients nonzero and yield only
$e^{i\phi}\in\{\pm1,\pm i\}$.

For the first equation, suppose $a,c\neq0$ and write
$a=2^i a'$ and $c=2^\ell c'$ with $a',c'$ odd.
If $i\neq\ell$, factoring out the smaller power of
two again leaves an odd integer greater than one.
Thus $i=\ell$, and
\begin{equation}
    a'^2+c'^2=2^{k-2i}.
\end{equation}
The left-hand side is congruent to $2$ modulo $4$,
so $k-2i=1$ and $a'^2+c'^2=2$.
Therefore $a',c'\in\{\pm1\}$, giving
\begin{equation}
    e^{i\phi}=\frac{\pm1\pm i}{\sqrt{2}}.
\end{equation}
The equation $2(b^2+d^2)=2^k$ gives the same phases
by the same argument. If only one coefficient is
nonzero, the phase is again $\pm1$ or $\pm i$.

Combining all cases, the only possibilities are
\begin{equation}
    e^{i\phi}\in
    \left\{
        \pm1,\ \pm i,\
        \frac{\pm1\pm i}{\sqrt{2}}
    \right\}.
\end{equation}
Each of these values belongs to
$\mathbb{Z}[1/\sqrt{2},i]$, so the allowed phases
are precisely
\begin{equation}
    \phi\equiv\frac{m\pi}{4}\pmod{2\pi},
    \qquad m=0,\ldots,7.
\end{equation}

\section{Symmetry-preserving simulation of scalar spin chirality}
\label{sec:chirality}

For three spin-$1/2$ particles, the scalar spin-chirality operator is
\begin{equation}
    \chi
    =
    \mathbf{S}_1\cdot
    \left(\mathbf{S}_2\times\mathbf{S}_3\right).
\end{equation}
Using $\mathbf{S}_j=\boldsymbol{\sigma}_j/2$, its Pauli expansion is
\begin{align}
    \chi
    =
    \frac{1}{8}
    (
        XYZ+YZX+ZXY
        \\-XZY-ZYX-YXZ
    ),
    \label{eq:chirality-pauli}
\end{align}
where, for example, $XYZ\equiv X_1Y_2Z_3$. Unlike the
two-site Heisenberg interaction, the Pauli strings in
Eq.~\eqref{eq:chirality-pauli} do not mutually commute. A direct
decomposition into independently implemented Pauli rotations would
therefore introduce a nontrivial product-formula error.

The chirality operator has a useful representation in terms of
two-site swap operators. For spin-$1/2$ particles,
\begin{equation}
    P_{ij}
    =
    \frac{1}{2}
    \left(
        I+\boldsymbol{\sigma}_i\cdot\boldsymbol{\sigma}_j
    \right).
\end{equation}
Using the Pauli commutation relations gives
\begin{equation}
    [P_{12},P_{23}]
    =
    -4i\chi,
\end{equation}
or equivalently
\begin{equation}
    \chi
    =
    \frac{i}{4}[P_{12},P_{23}].
    \label{eq:chirality-swap-commutator}
\end{equation}

Equation~\eqref{eq:chirality-swap-commutator} suggests a
group-commutator construction. Define
\begin{equation}
    K(\delta)
    =
    e^{-i\delta P_{12}}
    e^{ i\delta P_{23}}
    e^{ i\delta P_{12}}
    e^{-i\delta P_{23}}.
    \label{eq:chirality-group-commutator}
\end{equation}
Writing
\begin{equation}
    A=-i\delta P_{12},
    \qquad
    B=i\delta P_{23},
\end{equation}
the Baker--Campbell--Hausdorff expansion of
$e^Ae^Be^{-A}e^{-B}$ gives
\begin{equation}
    \log K(\delta)
    =
    [A,B]
    +\frac{1}{2}[A+B,[A,B]]
    +O(\delta^4).
\end{equation}
The leading commutator is
\begin{equation}
    [A,B]
    =
    \delta^2[P_{12},P_{23}]
    =
    -4i\delta^2\chi.
\end{equation}
Consequently,
\begin{equation}
    \log K(\delta)
    =
    -4i\delta^2\chi
    +\delta^3\mathcal{C}_3
    +O(\delta^4),
    \label{eq:chirality-K-expansion}
\end{equation}
where $\mathcal{C}_3$ denotes the third-order BCH contribution.

Under $\delta\mapsto-\delta$, the second-order term in
Eq.~\eqref{eq:chirality-K-expansion} is unchanged, while the
third-order term changes sign. We therefore define the symmetrized
sequence
\begin{equation}
    K_{\mathrm{sym}}(\delta)
    =
    K(\delta)K(-\delta).
    \label{eq:chirality-symmetric-sequence}
\end{equation}
Combining the two BCH expansions yields
\begin{equation}
    K_{\mathrm{sym}}(\delta)
    =
    \exp\!\left[
        -8i\delta^2\chi
        +O(\delta^4)
    \right].
    \label{eq:chirality-symmetric-expansion}
\end{equation}
Thus, the symmetrization cancels the third-order error while adding
the desired second-order contributions.

To approximate the chirality evolution $e^{-it\chi}$ for $t\geq0$,
choose an integer $r\geq1$ and set
\begin{equation}
    \delta
    =
    \sqrt{\frac{t}{8r}}.
\end{equation}
Repeating the symmetrized sequence $r$ times then gives
\begin{align}
    \left[
        K_{\mathrm{sym}}
        \left(
            \sqrt{\frac{t}{8r}}
        \right)
    \right]^r
    &=
    \exp\!\left[
        -it\chi
        +O\!\left(r\delta^4\right)
    \right]
    \nonumber\\
    &=
    \exp\!\left[
        -it\chi
        +O\!\left(\frac{t^2}{r}\right)
    \right].
    \label{eq:chirality-time-evolution}
\end{align}
Evolution for negative $t$ can be obtained by reversing the
orientation of the construction, or equivalently by using the
adjoint sequence. Each application of $K_{\mathrm{sym}}$ contains
eight fractional-swap gates, so the complete construction uses
$8r$ such gates.

Crucially, each constituent gate preserves the global spin symmetry.
For
\begin{equation}
    J_\alpha
    =
    \frac{1}{2}
    \sum_{k=1}^{3}\sigma_k^\alpha,
    \qquad
    \alpha\in\{x,y,z\},
\end{equation}
the swap operators satisfy
\begin{equation}
    [P_{ij},J_\alpha]=0.
\end{equation}
It follows that
\begin{equation}
    \left[
        K_{\mathrm{sym}}(\delta),
        J_\alpha
    \right]
    =0
\end{equation}
for every value of $\delta$, independent of the finite-$r$
approximation error in Eq.~\eqref{eq:chirality-time-evolution}.
The finite sequence therefore approximates the desired chirality
dynamics while preserving global $SU(2)$ symmetry exactly at every
stage.

The same conclusion holds when a fractional swap is replaced by a
symmetry-preserving approximant of the form
$e^{\pm i\delta'P_{ij}}$. Such synthesis errors change the implemented
exchange angles and hence the dynamical accuracy, but they do not
introduce leakage between total-spin sectors.

\section{BRS 2015  method}
In this section, we summarize the BRS 2015 RUS synthesis method for implementing a controlled-phase gate with over- and under-rotations about a fixed axis. We
restate the construction from Ref.~\cite{Bocharov2015RUS} which gives an axial $R_z$ gate. 
\\
Consider a unitary $V$ of the form
\begin{align}
    V=\frac{1}{\sqrt{2}^l}\begin{bmatrix}
        z & y\\
        -y^* & z^*
    \end{bmatrix}
\end{align}
where $z,y\in \mathbb{Z}[\omega]$ and $\omega = e^{i\pi/4}$, and $z,y$ satisfy the norm equation $|z|^2 + |y|^2 = 2^l$. Given an angle $\theta$, we can find a unitary $V$ such that $z^*/z $ approximates$ e^{i\theta}$. The unitary $V$ can be implemented using Clifford+T gates,
and the estimated T-count can be found as $t=2 \text{cde}$.  To approximate a cyclotomic rational $z^*/z$ to within $\epsilon$, a PSLQ integer relation algorithm can be used, by expressing $z=a\omega^3+b\omega^2+c\omega+d$ where $a,b,c,d\in \mathbb{Z}$, and $a,b,c,d$ are selected such that probability of success is maximized, and $\epsilon$ is minimized. 
We have the unitary $J(V)$ as 
\begin{align}
    J(V) = \begin{bmatrix}
        V & 0\\
        0 & V^{\dagger}
    \end{bmatrix}
\end{align}
Suppose the data qubit is in the state $\ket{\psi}=\alpha \ket{0} + \beta \ket{1}$ and the ancilla qubit is in the state $\ket{0}_a$. The action of $J(V)$ on the combined state is given by:
\begin{equation}
J(V)\bigl(\ket{\psi}_d\ket{0}_a\bigr)
=
\frac{1}{\sqrt{2^l}}
\Biggl[
\begin{aligned}
&\bigl(\alpha z\ket{0}_d+\beta z^*\ket{1}_d\bigr)\ket{0}_a\\
&\quad
+y^*\bigl(-\alpha\ket{0}_d+\beta\ket{1}_d\bigr)\ket{1}_a
\end{aligned}
\Biggr].
\end{equation}%

The success probability is independent of the input state and is given by
\begin{equation}
    p_{\mathrm{BRS}}
    =
    \frac{|z|^2}{2^l}.
\end{equation}
Conditioned on success, the normalized data state is
\begin{equation}
    \ket{\psi_{\mathrm{succ}}}
    =
    \frac{z}{|z|}
    \left(
        \alpha\ket{0}
        +
        \beta\frac{z^*}{z}\ket{1}
    \right).
\end{equation}
The factor \(z/|z|\) is a global phase. A failed attempt produces the
original state up to a Pauli-\(Z\) correction, which is Clifford, and the
protocol can therefore be repeated.

In the resource convention used in our numerical comparisons, synthesizing
\(V\) has estimated \(T\)-count \(2\operatorname{cde}(V)\), while implementing
the complete two qubit operation \(J(V)\) adds an overhead of nine \(T\)
gates \cite{Bocharov2015RUS}. Thus, the expected cost of one successfully completed axial rotation is
\begin{equation}
    \overline{T}_{R_z}^{\mathrm{BRS}}
    =
    \frac{2\operatorname{cde}(V)+9}
         {p_{\mathrm{BRS}}}.
\end{equation}
Our controlled-phase decomposition contains three such rotations. Assuming
that each rotation is repeated individually until it succeeds, its expected
total cost is therefore
\begin{equation}
    \overline{T}_{\cphase}^{\mathrm{BRS}}
    =
    \frac{3\left(2\operatorname{cde}(V)+9\right)}
         {p_{\mathrm{BRS}}}.
    \label{eq:brs-expected-cphase-t-count}
\end{equation}

\section{Implementation details of \texorpdfstring{$\cphase$}{CPHASE} compilation}\label{section: implementation}
In the main text we compare three methods: default gridsynth, BRS 2015 and LCU symmetry preserving methods. Since each of these compilations are approximate, they all incur an approximation error $\epsilon$, but it shows up in different ways. 
For the \textit{gridsynth} method, the errors do not necessarily act axially. From \cite{Selinger2013TDepth}, we have 
\begin{align}
    |\tilde{R_z}(\phi') - R_z(\phi)|\leq \epsilon
\end{align}
A $\cphase$ gate contains three such rotations. Expanding the difference between the ideal and compiled products and using unitary invariance of the operator norm gives the worst-case bound 
\begin{equation}
    \begin{split}
    \left\|\widetilde{\operatorname{CPHASE}}(\theta)
    -\operatorname{CPHASE}(\theta)\right\|
    &\leq\sum_{k=1}^{3}
    \left\|\widetilde R_{z,k}-R_{z,k}\right\|\\
    &\leq3\epsilon_{\mathrm{def}}.
    \end{split}
    \label{eq:gridsynth-cphase-bound}
\end{equation}
The surrounding Clifford gates preserve this bound for the compiled exchange gate. Equation~\eqref{eq:gridsynth-cphase-bound} is thus an upper bound on the total gate error;

For the BRS 2015 method, write
\begin{equation}
    \frac{z^*}{z}
    =
    e^{i(\theta/2+\epsilon)},
\end{equation}
where $\epsilon$ is the signed angular error in the synthesized
half-angle. The corresponding single-qubit gate is
\begin{equation}
    D
    =
    \begin{bmatrix}
        1&0\\
        0&z^*/z
    \end{bmatrix}.
\end{equation}
Up to a global phase, this is the $Z$ rotation
\begin{equation}
    R_z\left(\frac{\theta}{2}+\epsilon\right)
    =
    e^{-i(\theta+2\epsilon)Z/4}.
\end{equation}

Using the same synthesized rotation in the three-rotation
decomposition gives
\begin{align}
    \widetilde{\cphase}
    &=
    \operatorname{diag}
    \left(1,1,1,\left(\frac{z^*}{z}\right)^2\right)\\
    &=
    \cphase(\theta+2\epsilon).
\end{align}
Equivalently, the coherent error relative to the target gate can
be written as
\begin{equation}
    E_\epsilon
    =
    e^{-i\epsilon Z_1/2}
    e^{-i\epsilon Z_2/2}
    e^{i\epsilon Z_1Z_2/2}.
\end{equation}
Indeed,
\begin{equation}
    E_\epsilon
    =
    e^{-i\epsilon/2}\cphase(2\epsilon),
\end{equation}
and therefore
\begin{align}
    E_\epsilon
    e^{-i\theta/2}\cphase(\theta)
    =
    e^{-i(\theta+\epsilon)/2}
    \cphase(\theta+2\epsilon).
\end{align}
Thus, if
\begin{equation}
    \epsilon_{\mathrm{ang}}:=|\epsilon|,
\end{equation}
the CPHASE angular error is exactly
\begin{equation}
    \epsilon_{\mathrm{BRS}}=|(\theta+2\Delta)-\theta|
    =
    2\epsilon.
\end{equation}
The corresponding operator-norm error is
\begin{equation}
    \epsilon_{\mathrm{BRS}}^{op}
    =
    2\sin(\epsilon)
    \leq 2\epsilon.
\end{equation}

Finally, the LCU method has the following form. Define
\begin{equation}
    f := \frac{|z|^4}{|z|^4+|y|^4},
    \qquad
    \theta' := 2\arccos\sqrt{f}.
\end{equation}
The LCU candidate is selected to satisfy
\begin{equation}
    \left|\arccos\sqrt{f}-\frac{\theta}{2}\right|
    \leq \epsilon,
\end{equation}
where $\epsilon$ is the internal half-angle tolerance.
After removing the global phase, the implemented gate is
$\operatorname{CPHASE}(\theta')$. Its operator-norm error is
\begin{align}
    \epsilon_{\mathrm{LCU}}^{op}
    &:= \left\|
        \operatorname{CPHASE}(\phi')
        -\operatorname{CPHASE}(\phi)
    \right\| \\
    &= 2\left|\sin\left(\frac{\phi'-\phi}{2}\right)\right| \\
    &\leq 2\sin(\epsilon)
    \leq 2\epsilon,
\end{align}

while the angle error in the $\cphase$ gate is simply $\epsilon_{\mathrm{LCU}}=2\epsilon$

\end{document}